\documentclass[%
 reprint,
superscriptaddress,
 amsmath,amssymb,
 aps,prb,citeautoscript
]{revtex4-2}

\usepackage{graphicx}
\usepackage{dcolumn}
\usepackage{bm}
\usepackage{color}
\usepackage{hyperref}
\usepackage{tikz}
\usepackage{float}
\usepackage[figuresright]{rotating}
\usepackage{array}
\usepackage{tabularx}
\usepackage{comment}

\newcommand{\corr}[1]{{#1}}

\begin{document}

\title{Gelation dynamics of charged colloidal rods:\\critical behaviour and time-connectivity superposition principle}

\author{Lise Morlet-Decarnin}
    \affiliation{ENSL, CNRS, Laboratoire de Physique, F-69342 Lyon, France}
    
\author{Thibaut Divoux}
    \affiliation{ENSL, CNRS, Laboratoire de Physique, F-69342 Lyon, France}

\author{S\'ebastien Manneville}%
\email[]{sebastien.manneville@ens-lyon.fr      }
    \affiliation{ENSL, CNRS, Laboratoire de Physique, F-69342 Lyon, France}
\affiliation{Institut Universitaire de France}%

\date{\today}

\begin{abstract}
Charged colloidal particles can self-assemble into gel networks upon screening of electrostatic repulsion by added salt. While gelation of spherical colloids has been extensively studied, much less is known about the gelation dynamics of anisotropic colloids. Here, we focus on cellulose nanocrystals (CNCs) as prototypical rigid, highly charged rod-like colloids. In aqueous solution with salt, CNCs display a rich phase diagram ranging from gel at low solid content to glassy phases at higher concentrations. Building on our previous work [Morlet-Decarnin \textit{et al., ACS Macro Lett.}, 2023, \textbf{12}, 1733], we present an extensive study of the mechanical recovery dynamics of CNC suspensions following a strong shear. Time-resolved mechanical spectroscopy reveals a liquid-to-solid transition characterized by a well-defined critical gel point. The evolving viscoelastic spectra can be rescaled onto master curves, demonstrating a time-connectivity superposition principle and critical dynamics on both sides of the gel point. By varying the CNC weight fraction and salt concentration, we identify a boundary between gel and attractive glass states marked by clear changes in rheological observables, including the elastic and viscous moduli at the gel point and their high-frequency power-law exponents. Analysis of dynamic critical exponents and hyperscaling reveals pronounced asymmetry between pre-gel and post-gel dynamics and non-universal values of the dynamic exponent. These findings highlight gelation mechanisms specific to highly charged rod-like colloids and call for complementary microstructural characterization and theoretical modeling.
\end{abstract}

\maketitle

\section{Introduction}

A gel is a soft solid formed through the percolation of mesoscopic constituents into a space-spanning network. Gel constituents can be either polymers or colloids, i.e., nanometric to micrometric Brownian particles, which display attractive interactions.\cite{Cao:2020} The microstructure of the resulting network is responsible for the linear viscoelastic response of the gel. Understanding the formation dynamics of this network, referred to as the gelation process, is of great interest for the design of soft materials involving a gel-based matrix or solid materials obtained from soft precursors.\cite{Ioannidou:2016,Gebauer:2019,Fournier:2024} 
In polymeric systems undergoing gelation through chemical bonding, it is well established that the gelation dynamics go through a critical point, \cite{Winter:1986,Winter:1987,martin:1990,Scanlan:1991} also called the gel point, corresponding to the point when particles are arranged into a fractal network spanning over the whole sample volume with a specific self-similar hierarchical structure referred to as the ``critical gel''.\cite{forrest:1979,lin:1989} 
A similar phenomenology has been reported for physically-bonded macromolecules, \cite{Larsen:2008,Larsen:2008b,sun:2018,Suman:2020} where particles self-assemble under the combined influence of thermal noise and short-range interactions. 

In the case of colloidal gels, the relevance of the ``critical gel'' picture has been firmly established for systems composed of spherical particles,\cite{Ponton:2000,Ponton:2002,Wu:2013,Negi:2014,Keshavarz:2021} where short-range attractions lead to the formation of fractal, percolated networks with well-defined critical exponents. For \textit{anisotropic} colloidal particles, however, the situation is markedly more complex. A growing body of work has demonstrated that particle anisotropy strongly impacts aggregation pathways, network topology, and mechanical response, through directional interactions, steric constraints, and the coupling between translational and rotational degrees of freedom.\cite{Wilkins:2009,Han:2015,Kazem:2015,Melaugh:2025} Rod-like, plate-like, or patchy particles have been shown to form gels, arrested networks, or glassy states whose microstructure and rheology can differ from those of their spherical counterparts.\cite{Wierenga:1998,Huang:2009,solomon:2010,Krishna:2012,Thareja:2013,Wang:2019,Murphy:2020,Sudreau:2022b}

While percolated networks and gel-like states have been reported for a variety of anisotropic colloids, it remains unclear to what extent the critical gel framework developed for polymeric and spherical colloidal gels can be generalized to anisotropic colloids. In particular, time-resolved rheology has confirmed rigidity percolation together with time-cure superposition of the viscoelastic spectra in only one isolated case, namely, single-wall carbon nanotube suspensions.\cite{chen:2010} Sill, the existence and robustness of a well-defined gel point, the validity of time-connectivity superposition, and the applicability of hyperscaling relations in systems where anisotropy promotes heterogeneous connectivity and competes with glassy arrest, have not been systematically explored.

From an experimental point of view, gelation dynamics were first studied in polymer gels by measuring the viscosity before the gel point under continuous shear, and the relaxation modulus after the gel point through oscillatory measurements.\cite{Peniche:1974,adam:1979,gauthier:1980} The gel point was then determined as the point where the viscosity diverges, and an elastic modulus simultaneously emerges. However, although this approach provided the first evidence for criticality during gelation, it requires prior knowledge of when to switch from viscosity to relaxation modulus measurement. Moreover, shearing the material may interfere with its gelation dynamics.\cite{Hanley:1999,Drabarek:2002, Mokhtari:2008,Altmann:2004,Nelson:2022,Sudreau:2022}

Viscoelastic measurements under small-amplitude oscillatory shear provide a much less intrusive way to probe gel formation. In particular, time-resolved mechanical spectroscopy allows for an unambiguous determination of the critical point.
\cite{Mours:1994,Scanlan:1991,filippone:2015,kruse:2016,Geri:2018,arrigo:2020,Suman:2020}. Indeed, by probing the viscoelastic moduli at different frequencies as a function of time with sufficient resolution, one can identify the gel point as the point where the elastic and the viscous moduli have the same dependence on frequency $f$, or equivalently on the angular frequency $\omega=2\pi f$. By collapsing the resulting time-dependent viscoelastic spectra onto a master curve, a time-connectivity superposition principle, also referred to as time-aging superposition principle in soft glasses\cite{Joshi:2025},  and time-cure superposition principle in the polymer literature,\cite{Winter:1986,Chambon:1987,Martin:1988,Adolf:1990} has been established and studied in detail. Models have been devised to fit these master curves, and theories have been developed to predict the various dynamical exponents extracted from the relaxation spectra, referred to as critical exponents.\cite{Suman:2021} 
Furthermore, ``hyperscaling'' relations, i.e., relationships between critical exponents, have also been established and verified in many different {\it polymeric} systems, for both chemical and physical gels.\cite{Stauffer:1982,Winter:1987,Axelos:1990,Hodgson:1990,Winter:1997,Negi:2014,Suman:2020,Rouwhorst:2020}

The aim of the present article is to explore the gelation dynamics of {\it colloidal} suspensions made of {\it anisotropic} particles, namely cellulose nanocrystals (CNCs), in the presence of salt. Using time-resolved mechanical spectroscopy, we provide evidence for both the existence of a well-defined gel point and for the validity of time-connectivity superposition together with selected hyperscaling relations, in this specific colloidal system.

CNCs are solid rod-like colloidal particles extracted from natural sources such as wood, cotton, tunicates, and some bacteria. These nanocrystals exhibit a large geometrical anisotropy, with an aspect ratio varying between 3 and 30 depending on their source.\cite{Trache:2020,Habibi:2010,Lahiji:2010,Yucel:2021} When extracted using sulfuric acid hydrolysis, CNCs carry negative charges on their surface due to the presence of sulfate groups. \corr{Owing to electrostatic repulsion, CNCs dispersed in pure water form stable suspensions that display liquid crystal properties, including isotropic-to-nematic transition and cholesteric phases \cite{Revol:1992,Dong:1996,Xu:2024}. In the presence of salt, however,} screening of the repulsive forces leads to aggregation of the CNCs. Depending on the relative amounts of CNCs and salt in the suspension, different phases are formed, ranging from isotropic liquids at low CNC concentration, to gels (liquid crystals, resp.) at intermediate CNC concentration and high (low resp.) salt content, and to attractive glasses (repulsive glasses resp.) at high CNC concentration and high (low resp.) salt content. \cite{Xu:2018,Urenabenavides:2011,Xu:2020,Moud:2020,Qu:2021,Wojno:2022,Wojno:2023,Shim:2025}

In previous works, we have shown that, when a CNC gel is fluidized through the application of a strong shear flow, the viscoelastic solid reforms spontaneously upon flow cessation, and that the recovery dynamics follow both time-composition and time-connectivity superposition principles.\cite{Morlet-Decarnin:2022,morlet-decarnin:2023} More precisely, as detailed in Ref.~\onlinecite{Morlet-Decarnin:2022}, the temporal evolution of the elastic modulus can be rescaled onto a master curve using a characteristic time scale $t^*$, for all salt concentrations within the gel region of the CNC phase diagram. Such a {\it time-composition} superposition principle suggests that the salt content primarily impacts the gel recovery kinetics, rather than the gel microstructure. In Ref.~\onlinecite{morlet-decarnin:2023}, we have explored in more detail the recovery dynamics of CNC gels using time-resolved mechanical spectroscopy. Strikingly, our experiments have shown that CNC suspensions form a critical gel during their recovery, which follows a {\it time-connectivity} superposition principle similar to the one established for polymer gels. The time scales involved in the gelation process span more than 10 orders of magnitude, with $t^*$ reaching up to $10^6$~s. Such ultraslow dynamics question the relevance of phase diagrams determined at a given point in time, as already suggested in Ref.~\onlinecite{Xu:2018}. These results also highlight the key importance of systematically exploring the influence of both CNC and salt concentrations on the recovery dynamics, as well as the robustness of critical behaviors across the phase diagram.

In the present article, building upon Ref.~\onlinecite{morlet-decarnin:2023}, we use time-resolved mechanical spectroscopy to show that the time-connectivity superposition principle remains valid for a wide range of CNC and salt concentrations, and that CNC suspensions consistently go through a critical point during their recovery. Furthermore, our dynamical characterization of the gel point suggests a way to distinguish between the gel phase and the attractive glass phase across the CNC--salt phase diagram. This paper is organized as follows. Section~\ref{section2} presents the experimental system and methods, as well as data analysis. In Sect.~\ref{section 3}, we examine the recovery dynamics of a suspension at a given CNC and salt concentration and introduce the time-connectivity superposition principle as in Ref.~\onlinecite{morlet-decarnin:2023}. The influence of the CNC concentration on the recovery dynamics is investigated and discussed in Sect.~\ref{section 4}, while Sect.~\ref{section 5} deepens this analysis by exploring the effect of the ionic strength. Finally, in Sect.~\ref{section 6}, we propose some interpretations and provide a list of open questions raised by the present results.

\begin{figure}[!t]
    \centering
    \includegraphics[width=0.45\textwidth]{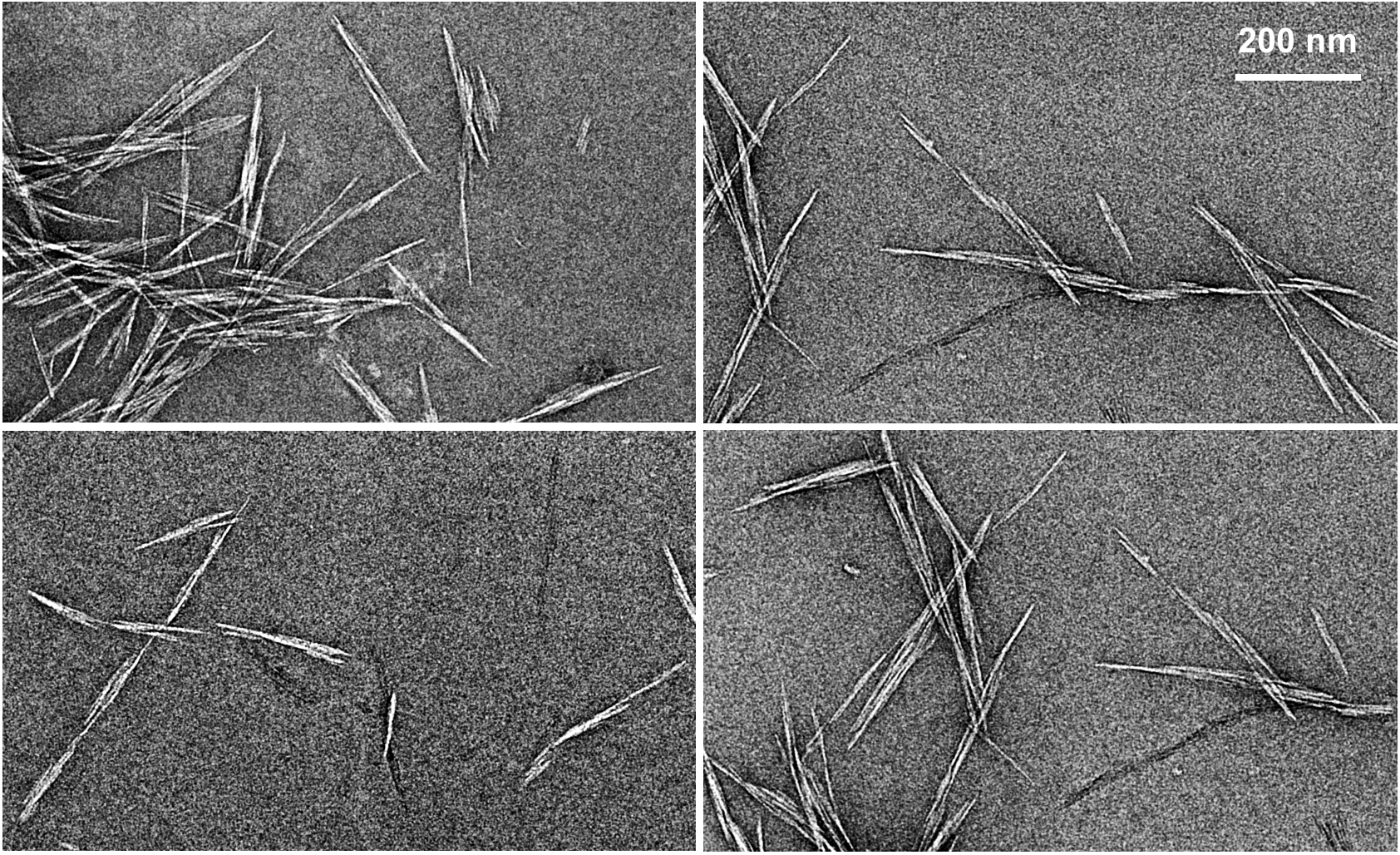}
    \caption{Transmission electron microscopy images of a dilution of the commercial CNC suspension from Celluforce (courtesy of Bruno Jean and Jean-Luc Puteaux). The 6.4~wt\% CNC suspension was exposed to mechanical stirring during 5~min at 2070 rpm (IKA RW 20 Digital mixer equipped with an R1402 blade dissolver) prior to dilution.}
    \label{fig:TEM}
\end{figure}

\begin{figure*}[htbp]
    \centering
    \includegraphics[width=0.9\textwidth]{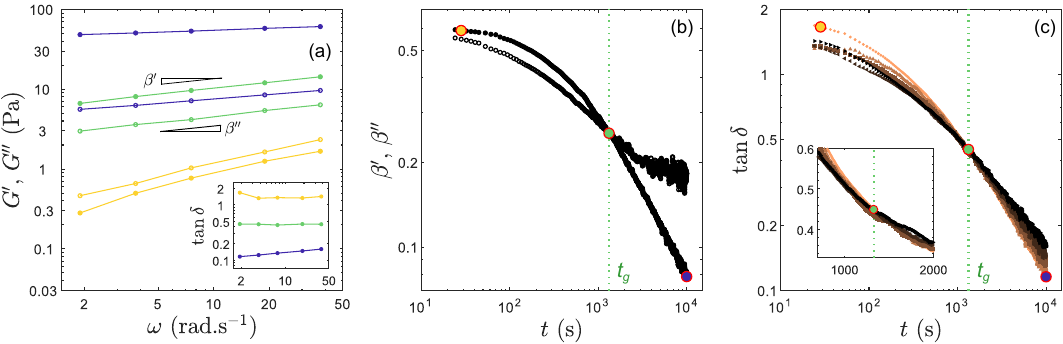}
    \caption{Time-resolved mechanical spectroscopy during the recovery of a $3.2$~wt\% CNC suspension containing $12$~mM NaCl. (a)~Elastic modulus $G'$ ($\bullet$) and viscous modulus $G''$ ($\circ$) as a function of angular frequency $\omega$ at different times across the sol-gel transition ($t=28$, 1330, and 10000~s, from yellow to dark blue). Inset: corresponding loss factor $\tan \delta=G''/G'$ as a function of $\omega$. The data shown in the inset correspond to the time points highlighted with larger colored symbols in (b) and (c). (b)~Temporal evolution of the frequency-scaling exponents $\beta'$ ($\bullet$) and $\beta''$ ($\circ$), extracted from power-law fits of $G'(\omega,t)$ and $G''(\omega,t)$ respectively. (c)~Temporal evolution of the loss factor $\tan \delta (t)$ measured at the five frequencies of the multiwave signal: $\omega/(2\pi)=0.3$~Hz ($\bullet$), 0.6~Hz ($\blacktriangle$), 1.2~Hz ($\blacksquare$), 3~Hz ($\blacktriangleleft$), and 6~Hz ($\blacktriangleright$). Inset: same data displayed in linear scales close to the gel point. Darker colors correspond to higher frequencies. The vertical green dotted lines in (b) and (c) highlight the gelation time $t_g$, identified by the frequency-independent loss factor.}
    \label{fig:reprise_multiwave_tg}
\end{figure*}

\section{Material and Methods}
\label{section2}

\subsection{Samples}
\label{mat_meth_samples}

CNC suspensions are prepared using a commercial aqueous suspension of CNC extracted from wood (CelluForce). These CNC particles have a typical length of about 120~nm, and a typical diameter of about 10~nm, with standard deviations of 40~nm and 2~nm, respectively, as determined from the analysis of transmission electron microscopy images. As shown in Fig.~\ref{fig:TEM}, they are present both as individual nanorods of diameter 3--4~nm and as bundles composed of nanorods arranged parallel to each other, or even into lines of a few CNCs assembled end-to-end. Measurements performed on the commercial suspension yielded a zeta-potential ranging between $-36.0$~mV and $-41.7$~mV, and the surface charge was estimated to be about $0.2e.\text{nm}^{-2}$.\cite{Note_BJean} Such estimates are consistent with measurements on other sulfated CNCs under similar conditions.\cite{Cherhal:2015,Prathapan:2016}

The CNC weight fraction in the commercial suspension is 6.4~wt\%. This suspension is diluted by adding an aqueous solution of NaCl (Merck) under high shear, following the procedure described in Ref.~\onlinecite{morlet-decarnin:2023}. The resulting samples contain a final CNC weight fraction $w_\mathrm{CNC}$ ranging from 0.75 to 5.5~wt\%, and a NaCl concentration between 5 and 22~mM. According to Refs.~\onlinecite{Xu:2018,Xu:2020}, these compositions fall within the gel or attractive glass regions of the CNC phase diagram.

\subsection{Rheological setup}
\label{mat_meth_rheo}

The rheological properties of the CNC suspensions are measured using a stress-controlled rheometer (AR-G2, TA Instruments) equipped with a smooth cylindrical Taylor-Couette geometry (height 58~mm, inner rotating cylinder of radius 24~mm, outer fixed cylinder of radius 25~mm, and gap $e=1$~mm) made of poly(methyl methacrylate). In order to prevent evaporation, the cell is closed with a homemade lid. The temperature is controlled at $T=23.0\pm 0.1$~$^{\circ}$C, thanks to a water circulation around the cell. 

\subsection{Time-resolved mechanical spectroscopy}
\label{mat_meth_protocol}

After loading the sample into the rheometer cell, it is fully fluidized by applying a high shear rate $\dot{\gamma}=1000$~s$^{-1}$ during 60~s, ensuring a reproducible initial state. When shear is stopped, which defines the origin of time $t=0$~s, the sample is in a liquid-like state and progressively recovers solid-like properties as CNC particles aggregate and eventually percolate to form a space-spanning network. To characterize the dynamics of the sol-gel transition during recovery at rest, we measure the time-evolution of the viscoelastic properties of the sample at different frequencies using time-resolved mechanical spectroscopy.\cite{Mours:1994} 

Specifically, we apply a ``multiwave signal'' that allows us to measure the viscoelastic properties of our sample at 5 frequencies simultaneously, every few seconds, over durations ranging from 5000~s to 50000~s. The input signal is a small-amplitude periodic torque or strain signal, depending on the sample, composed of the sum of five sinusoidal signals of different frequencies and amplitudes. A Fourier transform of the resulting strain or torque signal, directly performed by the rheometer software, yields the elastic modulus $G'$, the viscous modulus $G''$, and the loss factor $\tan \delta=G''/G'$ at the five different frequencies of interest simultaneously, every $\Delta t_\mathrm{exp}$. Two measurement protocols are used depending on the CNC weight fraction.
(i)~For samples containing less than 3.2~wt\% CNC, the torque is imposed with constant amplitudes of $5,\ 4,\ 3,\ 2,\ 1$~$\mu$N.m at frequencies $\omega/(2\pi)=0.3,\ 0.6,\ 0.9,\ 1.2,\ 1.5$~Hz, respectively, and with $\Delta t_\mathrm{exp}=5$~s. 
(ii)~For samples containing 3.2~wt\% CNC or more, the strain is imposed with a constant amplitude chosen between 0.2 to 2~\% at frequencies $\omega/(2\pi)=0.3,\ 0.6,\ 1.2,\ 3,\ 6$~Hz, and with $\Delta t_\mathrm{exp}=8$~s.
In both cases, the torque or strain amplitudes are chosen so that the maximum strain induced from the entire multiwave signal remains within the linear viscoelastic regime.
Moreover, the selected range of frequencies ensures that (i)~the sample properties do not evolve significantly during a single measurement step, i.e., the mutation number $N_{mu}$ remains small:\cite{Winter:1988,Mours:1994} $N_{\rm mu}=(\Delta t_{\rm exp}/G')\,(\partial G'/\partial t)\ll 1$, and (ii)~inertial effects are negligible at large frequencies. We checked that torque-controlled and strain-controlled protocols yield identical results for a 3.2~wt\% CNC suspension containing 12~mM of NaCl.

\subsection{Data analysis}
\label{mat_meth_data_anal}

At each measurement time, the time-resolved elastic and viscous moduli are fitted by power-law behaviors of the angular frequency $\omega$, $G'(\omega,t)\sim\omega^{\beta'(t)}$ and $G''(\omega,t)\sim\omega^{\beta''(t)}$, respectively, as illustrated in Fig.~\ref{fig:reprise_multiwave_tg}(a) for a typical sample with $w_\mathrm{CNC}=3.2$~wt\% and [NaCl]=12~mM. 
The corresponding exponents $\beta'$ and $\beta ''$ are plotted as a function of time in Fig.~\ref{fig:reprise_multiwave_tg}(b). Following previous works, \cite{Winter:1986,Winter:1987,Suman:2020} the \textit{gelation time} $t_g$ is defined as the time at which the elastic and viscous moduli exhibit the same power-law dependence on frequency, i.e., $\beta'(t_g)=\beta''(t_g)$ (see green point in Fig.~\ref{fig:reprise_multiwave_tg}). The corresponding exponent, common to both $G'$ and $G''$, defines the \textit{critical exponent} $\beta$, and at this gelation time, the loss tangent $\tan \delta=G''/G'$ is independent of angular frequency, [see data in green in the inset of Fig.~\ref{fig:reprise_multiwave_tg}(a)]. Equivalently, the five curves $\tan \delta(t)$ corresponding to the five frequencies probed by the multiwave signal intersect at $t_g$, as confirmed in Fig.~\ref{fig:reprise_multiwave_tg}(c).

For all samples, the loss tangent and the viscoelastic spectra are further rescaled onto master curves similar to those of Figs.~\ref{fig:courbes_maitresses_3p2pc_CNC_12mM_NaCl} and \ref{fig:courbes_maitresses_var_CNC_concentration} using an open-source code developed by Lennon \textit{et al.}\cite{Lennon:2023} This code determines the optimal horizontal and vertical shift factors required to collapse the data onto a single master curve by maximizing their overlap.
First, the loss tangent spectra $\tan \delta (\omega)$ are horizontally shifted toward higher frequencies before the gel point (for $t<t_g$), and toward lower frequencies after the gel point (for $t>t_g$), using the first spectrum measured after shear cessation as a reference. The master curve obtained for the same $\tan \delta (\omega)$ data as in Fig.~\ref{fig:reprise_multiwave_tg}, yet plotted for more measurement times in Fig.~\ref{fig:courbes_maitresses_3p2pc_CNC_12mM_NaCl}(a), is shown in Fig.~\ref{fig:courbes_maitresses_3p2pc_CNC_12mM_NaCl}(c). The rescaled angular frequency $\tilde{\omega}$ is defined as $\tilde{\omega}=a(t)\omega$ where $a(t)$ is the time-dependent horizontal shift factor inferred from the rescaling code and displayed in Fig.~\ref{fig:courbes_maitresses_3p2pc_CNC_12mM_NaCl}(b).

\begin{figure}[!t]
    \centering
    \includegraphics[width=0.35\textwidth]{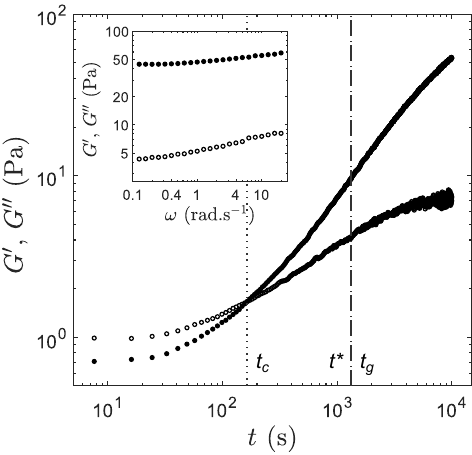}
    \caption{Temporal evolution of the elastic modulus $G'$ ($\bullet$) and viscous modulus $G''$ ($\circ$) measured at fixed frequency $\omega/(2\pi)=1.2$~Hz following a 60~s preshear at $\dot{\gamma}=1000$~s$^{-1}$. The vertical dotted line at $t_c=165$~s marks the crossover time at which $G'=G''$ at this frequency. The vertical dotted line at $t_g=1330$~s indicates the gelation time identified from time-resolved mechanical spectroscopy by the frequency-independent loss factor. The vertical dashed line at $t^*=1314$~s denotes the inflection time introduced in Ref.~\onlinecite{Morlet-Decarnin:2022}. Inset: Viscoelastic spectra $G'(\omega)$ ($\bullet$) and $G''(\omega)$ ($\circ$) measured after $10000$~s of recovery under small-amplitude oscillatory shear (strain of amplitude 0.2~\%) by sweeping down the angular frequency logarithmically from 20 to 0.1~rad.s$^{-1}$ within 1500~s. Experiment performed on the same 3.2~wt\% CNC suspension containing 12~mM of NaCl as in Fig.~\ref{fig:reprise_multiwave_tg}.}
    \label{fig:reprise_1freq}
\end{figure}

The same frequency shift is then applied to $G'(\omega)$ and $G''(\omega)$, which are further shifted vertically using the same code to construct a master curve for the viscoelastic moduli. For all samples, we find that the resulting time-dependent shift factors associated with $G'$ and $G''$, are very similar. Therefore, we use a single time-dependent vertical shift factor $b(t)$, defined as the average of the two shift factors and displayed in Fig.~\ref{fig:courbes_maitresses_3p2pc_CNC_12mM_NaCl}(e), to rescale both moduli such that $\tilde{G'}=b(t) G'$ and $\tilde{G''}=b(t) G''$, as illustrated in Fig.~\ref{fig:courbes_maitresses_3p2pc_CNC_12mM_NaCl}(f).

\begin{figure*}[!t]
    \centering
    \includegraphics[width=0.9\textwidth]{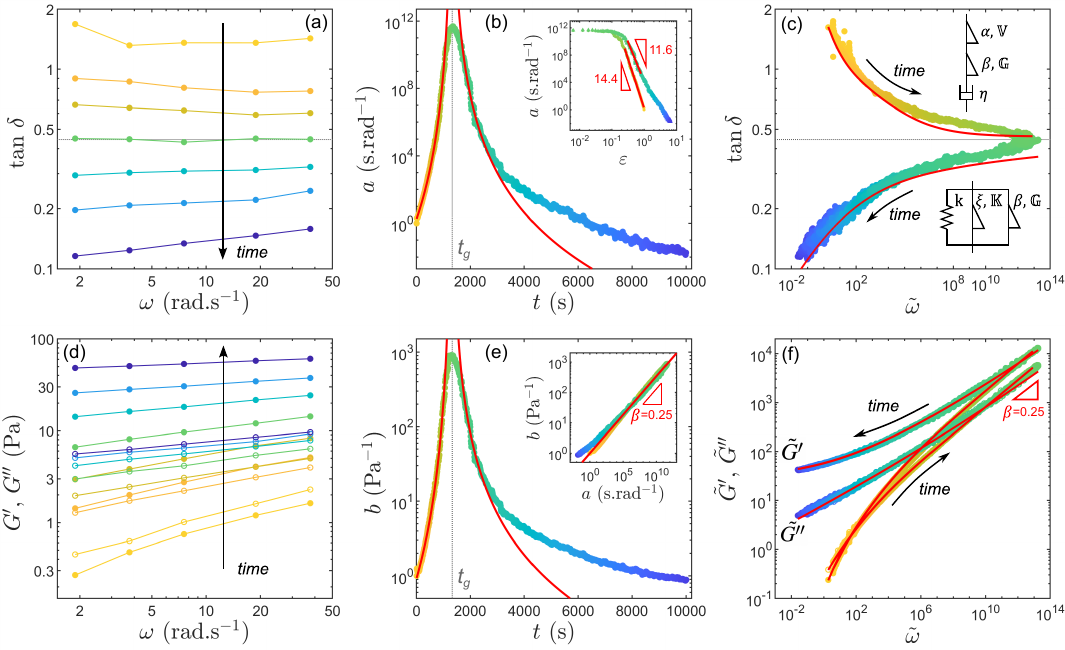}
    \caption{Time-resolved mechanical spectroscopy of the sol-gel transition in a $3.2$~wt\% CNC suspension containing $12$~mM of NaCl. (a)~Loss factor $\tan \delta$ as a function of angular frequency $\omega$ at different times after shear cessation, from $t=8$ up to $10000$~s (from yellow to dark blue). (b)~Temporal evolution of the horizontal shift factor $a(t)$, with the initial value arbitrarily set to $a(0)=1$~s.rad$^{-1}$. Red curves show the best power-law fits of the data in the vicinity of the gel point, $a\sim\varepsilon^{-y}$, where $\varepsilon=|t-t_g|/t_g$, yielding exponents $y_l=14.4\corr{\pm 0.8}$ for $t<t_g$ and $y_g=11.6\corr{\pm 0.7}$ for $t>t_g$. Inset: $a$ plotted as a function of $\varepsilon$ on logarithmic scales together with the corresponding fits, restricted to the fitting intervals. (c)~Master curve for $\tan \delta$ as a function of the rescaled angular frequency $\tilde \omega = a(t)\omega$. Red curves show the best fits obtained using a fractional Maxwell model for $t<t_g$ (upper curve and sketch) and a fractional Kelvin-Voigt model for $t>t_g$ (lower curve and sketch). (d)~Elastic modulus $G'$ ($\bullet$) and viscous modulus $G''$ ($\circ$) as a function of $\omega$ at the same times as in (a). (e)~Temporal evolution of the vertical shift factor $b(t)$, with $b(0)=1$~Pa$^{-1}$. Red curves show the best power-law fits of the data in the vicinity of the gel point, $b\sim\varepsilon^{-z}$, yielding exponents $z_l=3.9\corr{\pm 0.2}$ for $t<t_g$ and $z_g=2.7\corr{\pm 0.15}$ for $t>t_g$. Inset: $b(t)$ as a function of $a(t)$, with the red line indicating a power-law relation of exponent $\beta=0.25$. (f)~Master curve for the rescaled elastic modulus $\Tilde{G'}$ and viscous modulus $\Tilde{G''}$ as a function of $\tilde \omega = a(t)\omega$. Red curves show the best fits obtained using a fractional Maxwell model for $t<t_g$ (lower curves) and a fractional Kelvin-Voigt model for $t>t_g$ (upper curves). The dotted lines at $\tan \delta(t_g)=0.44$ in (a) and (c), and at $t_g=1330$~s in (b) and (e) indicate the gel point.}
    \label{fig:courbes_maitresses_3p2pc_CNC_12mM_NaCl}
\end{figure*}

\section{Recovery dynamics: critical-like gelation and time-connectivity superposition}
\label{section 3}

\subsection{Recovery dynamics at a single frequency}
\label{recovery_1HZ}

A first, simple approach to probe the recovery dynamics of a colloidal gel following shear cessation consists in monitoring its viscoelastic properties at a single frequency. Figure~\ref{fig:reprise_1freq} illustrates such a temporal evolution for a 3.2~wt\% CNC suspension containing 12~mM NaCl, extracted from time-resolved mechanical spectroscopy measurements at the frequency $\omega/(2\pi)=1.2$~Hz. Both the elastic modulus $G'$ and the viscous modulus $G''$ increase with time as the sample undergoes the sol-gel transition at rest after shear cessation. At short times, the viscous modulus $G''$ is larger than the elastic modulus $G'$, indicating a viscoelastic liquid state. However, $G''$ builds up more slowly than $G'$, so that the two moduli cross at a time $t_c=165$~s, referred to as the \textit{crossover time}, where $G'(t_c)=G''(t_c)$ (see the vertical dotted line in Fig.~\ref{fig:reprise_1freq}). At longer times, the elastic modulus overcomes the viscous modulus, pointing to a viscoelastic solid state.

The emergence of a solid-like state is confirmed by recording the full viscoelastic spectrum  at long times. As shown in the inset of Fig.~\ref{fig:reprise_1freq}, the elastic modulus is larger than the viscous modulus over the whole range of frequencies under study, including in the low-frequency limit. However, a simple single-frequency measurement, such as the determination of the crossover time $t_c$, does not allow one to fully identify the sol-gel transition nor to determine the corresponding gelation time $t_g$, defined as the time required for CNC particles to aggregate into a percolated, load-bearing network spanning over the whole sample volume. In particular, following on our previous work,\cite{Morlet-Decarnin:2022} we may introduce another characteristic time scale for the single-frequency recovery dynamics, namely the time $t^*$ at which the elastic modulus $G'(t)$ exhibits an inflection point in logarithmic scales (see the vertical dashed line indicating $t^*=1314$~s in Fig.~\ref{fig:reprise_1freq}). We show below that this \textit{inflection time} $t^*$, which is always much larger than $t_c$, actually corresponds to the gelation time $t_g$ introduced above in Sect.~\ref{mat_meth_data_anal}.

\subsection{Recovery dynamics through time-resolved mechanical spectroscopy}
\label{recovery_multiwave}

In order to properly identify the gel point and the corresponding gelation time, one needs to consider the time evolution of the sample viscoelastic properties measured at several frequencies. As recalled in Sect.~\ref{mat_meth_data_anal}, the ``true'' gel point is defined as the point where the loss tangent $\tan \delta$ is frequency independent, and both the elastic and the viscous moduli follow the same power-law dependence on angular frequency, characterized by a critical exponent $\beta$:\cite{Winter:1986,Winter:1987,Suman:2020} 
\begin{equation}
    G'(\omega) \sim G''(\omega) \sim \omega^{\beta}\,.
    \label{eq:gel_point}
\end{equation}

Figures~\ref{fig:courbes_maitresses_3p2pc_CNC_12mM_NaCl}(a) and ~\ref{fig:courbes_maitresses_3p2pc_CNC_12mM_NaCl}(d) show the time-resolved viscoelastic spectra measured during the recovery of a suspension containing 3.2~wt\% CNC and 12~mM NaCl, from $t=8$~s up to $t=10000$~s after shear cessation, using the multiwave protocol described in Section~\ref{mat_meth_protocol}. For all frequencies contained in the multiwave signal, both $G'$ and $G''$ increase with time. As already observed for a single frequency in Fig.~\ref{fig:reprise_1freq}, the elastic modulus increases faster than the viscous modulus, so that the loss factor $\tan \delta$ decreases over time [see also Fig.~\ref{fig:reprise_multiwave_tg}(c)]. Moreover, at all times, both moduli scale as power laws of the angular frequency $\omega$, yet with different exponents, with $\beta'>\beta''$ at short recovery times [see Fig.~\ref{fig:reprise_multiwave_tg}(b)]. Consequently, $\tan \delta$ initially decreases as a power-law of the angular frequency. 

As recovery proceeds, the loss tangent flattens until a point where it becomes independent of angular frequency, as highlighted by the horizontal dotted line at $\tan \delta(t_g)=0.44$ in Fig.~\ref{fig:courbes_maitresses_3p2pc_CNC_12mM_NaCl}(a). This behavior is characteristic of a critical gel, and the corresponding time $t_g=1330$~s defines the gelation time. Beyond the gel point, $\tan \delta$ increases as a power-law of the angular frequency, with an exponent that grows with time as $G'$ progressively tends toward a plateau as a function of $\omega$. 

This analysis shows that, upon recovery from a strong shear, the CNC suspension examined in Figs.~\ref{fig:reprise_multiwave_tg}--\ref{fig:courbes_maitresses_3p2pc_CNC_12mM_NaCl} undergoes a sol-gel transition at time $t_g=1330$~s, i.e., much later than the liquid-like to solid-like transition inferred from the single-frequency crossover at $t_c=165$~s in Sect.~\ref{recovery_1HZ}. 
The fact that $\tan \delta(t_g)=0.44$, significantly smaller than unity, confirms that the sample exhibits an elastic-like behavior at the gel point. Such a marked separation between the crossover time $t_c$ and the gelation time $t_g$ is particularly striking, since the crossover time is often taken as a proxy for the gelation time in the colloid literature\cite{Kamp:2009,Chen:2023,Smit:2025} and $t_c \simeq t_g$ has been reported for only a few colloidal systems.\cite{Ponton:2002, Negi:2014,Keshavarz:2021, Suman:2020}. Finally, within experimental uncertainty, the gelation time inferred from time-resolved mechanical spectroscopy [Fig.~\ref{fig:courbes_maitresses_3p2pc_CNC_12mM_NaCl}(a)] coincides with the inflection time $t^*=1314$~s identified in the single-frequency measurements (see Fig.~\ref{fig:reprise_1freq}), which fully confirms the interpretation of $t^*$ hypothesized in our previous work.\cite{Morlet-Decarnin:2022}.

\subsection{Time-connectivity superposition across the sol-gel transition}
\label{time_connectivity}

In order to collapse the time-resolved viscoelastic spectra of Figs.~\ref{fig:courbes_maitresses_3p2pc_CNC_12mM_NaCl}(a) and \ref{fig:courbes_maitresses_3p2pc_CNC_12mM_NaCl}(d) onto single master curves, we use the rescaling analysis introduced in Ref.~\onlinecite{morlet-decarnin:2023} and described above in Section~\ref{mat_meth_data_anal}. This procedure yields the time-dependent horizontal and vertical shift factors $a(t)$ and $b(t)$, respectively, displayed in Figs.~\ref{fig:courbes_maitresses_3p2pc_CNC_12mM_NaCl}(b) and \ref{fig:courbes_maitresses_3p2pc_CNC_12mM_NaCl}(e), respectively, as well as master curves for $\tan \delta$ and for the rescaled viscoelastic moduli $[\tilde{G'},\tilde{G''}]=[b(t)G', b(t)G'']$ plotted as a function of the rescaled angular frequency $\tilde{\omega}=a(t) \omega$ in Figs.~\ref{fig:courbes_maitresses_3p2pc_CNC_12mM_NaCl}(c) and \ref{fig:courbes_maitresses_3p2pc_CNC_12mM_NaCl}(f), respectively. Interestingly, close to the gel point, both $a(t)$ and $b(t)$ diverge as power laws of the time to gelation $|t-t_g|$ [see red curves in Figs.~\ref{fig:courbes_maitresses_3p2pc_CNC_12mM_NaCl}(b) and \ref{fig:courbes_maitresses_3p2pc_CNC_12mM_NaCl}(e)]. 
Such critical-like behavior has already been reported for the gelation dynamics of several other colloidal systems, including carbon nanotube dispersions,\cite{chen:2010} aluminosilicate and silica suspensions,\cite{Keshavarz:2021} Laponite suspensions,\cite{Suman:2020}, and polyamide rods.\cite{he:2021}
However, in the present system, the exponents characterizing the power-law divergence of $a(t)$ and $b(t)$ at $t_g$ differ significantly on either side of the gel point. \corr{We denote these exponents by $y_i$ and $z_i$, where $i=l$ for $t<t_g$ and $i=g$ for $t>t_g$, namely:
\begin{eqnarray}
    a(t) \sim (t_g-t)^{y_l} \hbox{\rm ~and~} b(t) \sim (t_g-t)^{z_l} \hbox{\rm ~for~} t<t_g\,,\\
a(t) \sim (t-t_g)^{y_g} \hbox{\rm ~and~} b(t) \sim (t-t_g)^{z_g}
     \hbox{\rm ~for~} t>t_g\,.
\end{eqnarray}}
For the sample investigated in Fig.~\ref{fig:courbes_maitresses_3p2pc_CNC_12mM_NaCl}, we report $y_l=14.4\corr{\pm 0.8}$ and $y_g=11.6\corr{\pm 0.7}$ for $a(t)$, and $z_l=3.9\corr{\pm 0.2}$ and $z_g=2.7\corr{\pm 0.15}$ for $b(t)$, \corr{so that $y_l > y_g$ and $z_l > z_g$, outside experimental uncertainty. In other words, the shift factors $a(t)$ and $b(t)$ both diverge more sharply in time at the gel point from the pre-gel state ($t<t_g$) than from the post-gel state ($t>t_g$).} Such a pronounced asymmetry with respect to the gel point contrasts with what has been observed or assumed in many polymeric and colloidal systems, where similar exponents are typically reported on both sides of the transition.\cite{Adolf:1990, Suman:2020, Larsen:2008, Scanlan:1991,he:2021}

\begin{sidewaysfigure*}
    \centering
    \includegraphics[width=0.8\textwidth]{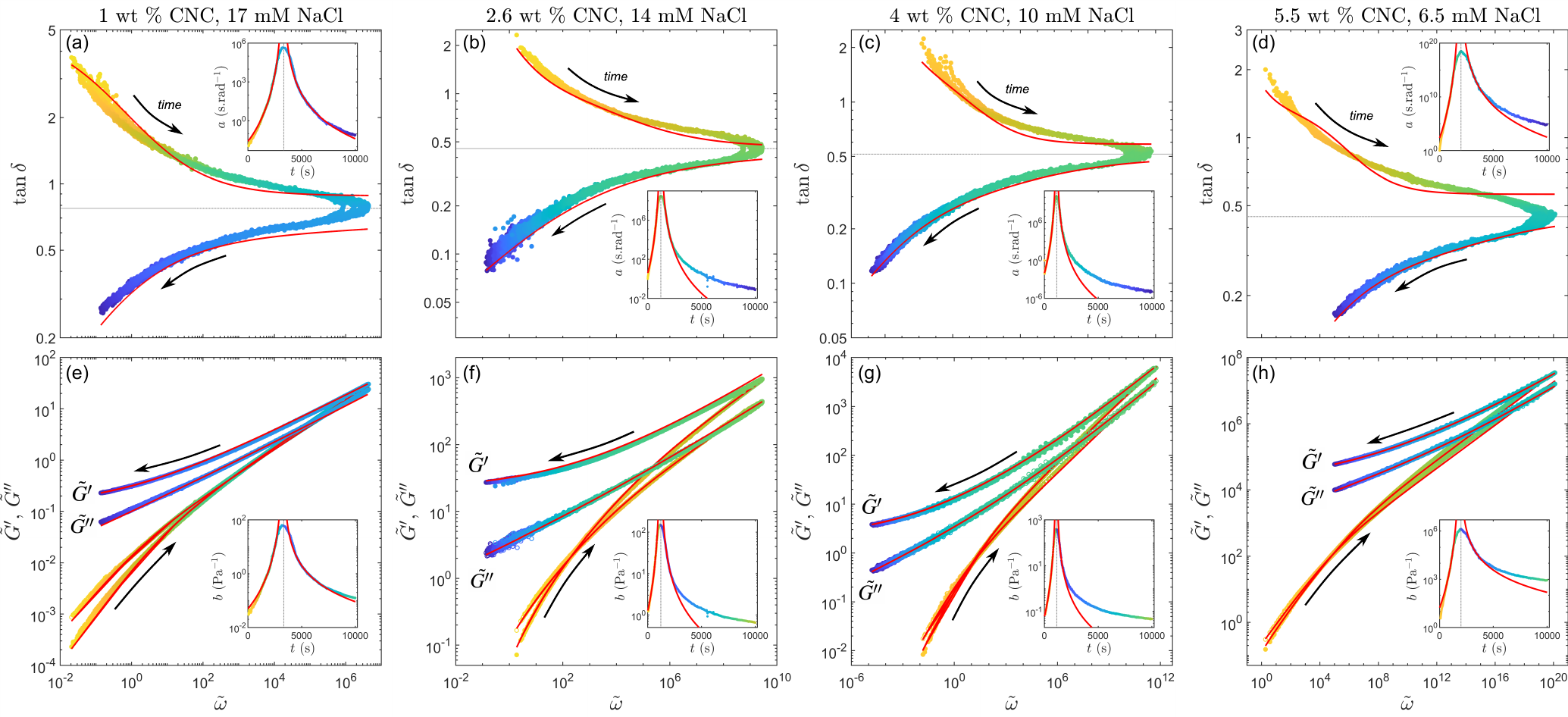}
    \caption{Time-resolved mechanical spectroscopy of the sol-gel transition in CNC suspensions with varying CNC and salt concentrations. From left to right: 1~wt\% CNC and 17~mM NaCl, 2.6~wt\% CNC and 14~mM NaCl, 4~wt\% CNC and 10~mM NaCl, and 5.5~wt\% CNC and 6.5~mM NaCl. (a)--(d)~Master curves for the loss factor $\tan \delta$ as a function of the rescaled angular frequency $\tilde \omega = a(t)\omega$. Insets: temporal evolution of the horizontal shift factor $a(t)$, with the initial value arbitrarily set to $a(0)=1$~s.rad$^{-1}$. (e)--(h)~Master curves for the rescaled elastic modulus $\tilde{G'}$ ($\bullet$) and viscous modulus $\tilde{G''}$ ($\circ$) as a function of $\tilde \omega = a(t)\omega$. Insets: temporal evolution of the vertical shift factor $b(t)$, with $b(0)=1$~Pa$^{-1}$. In all graphs, dotted lines indicate the gel point $t_g$. Red curves in the main graphs show the best fits of the data using a fractional Maxwell model for $t<t_g$ (upper curve) and a fractional Kelvin-Voigt model for $t>t_g$ (lower curve). Red curves in the insets show the best power-law fits in the vicinity of the gel point, $a\sim\varepsilon^{-y}$, with exponents $y_l$ for $t<t_g$ and $y_g$ for $t>t_g$, and $b\sim\varepsilon^{-z}$, with exponents $z_l$ for $t<t_g$ and $z_g$ for $t>t_g$, where $\varepsilon=|t-t_g|/t_g$.}
    \label{fig:courbes_maitresses_var_CNC_concentration}
\end{sidewaysfigure*}

Remarkably, the master curves in Fig.~\ref{fig:courbes_maitresses_3p2pc_CNC_12mM_NaCl}(f) span more than five orders of magnitude in rescaled viscoelastic moduli, and over sixteen orders of magnitude in rescaled angular frequency, on both sides of the gel point. This exceptionally broad dynamic range allows us to resolve both pre-gel and post-gel dynamics, whereas in most colloidal systems, gelation is too fast to access the dynamics before the gel point.\cite{chen:2010,Keshavarz:2021} To our knowledge, only two other colloidal systems, namely Laponite suspensions\cite{Suman:2020} and suspensions of polyamide rods\cite{he:2021}, have allowed for comparable pre-gel measurements. The successful construction of master curves on both sides of the gel point demonstrates the existence of a time-connectivity superposition principle that holds throughout the sol-gel transition.

\begin{figure*}[!t]
    \centering
    \includegraphics[width=0.75\textwidth]{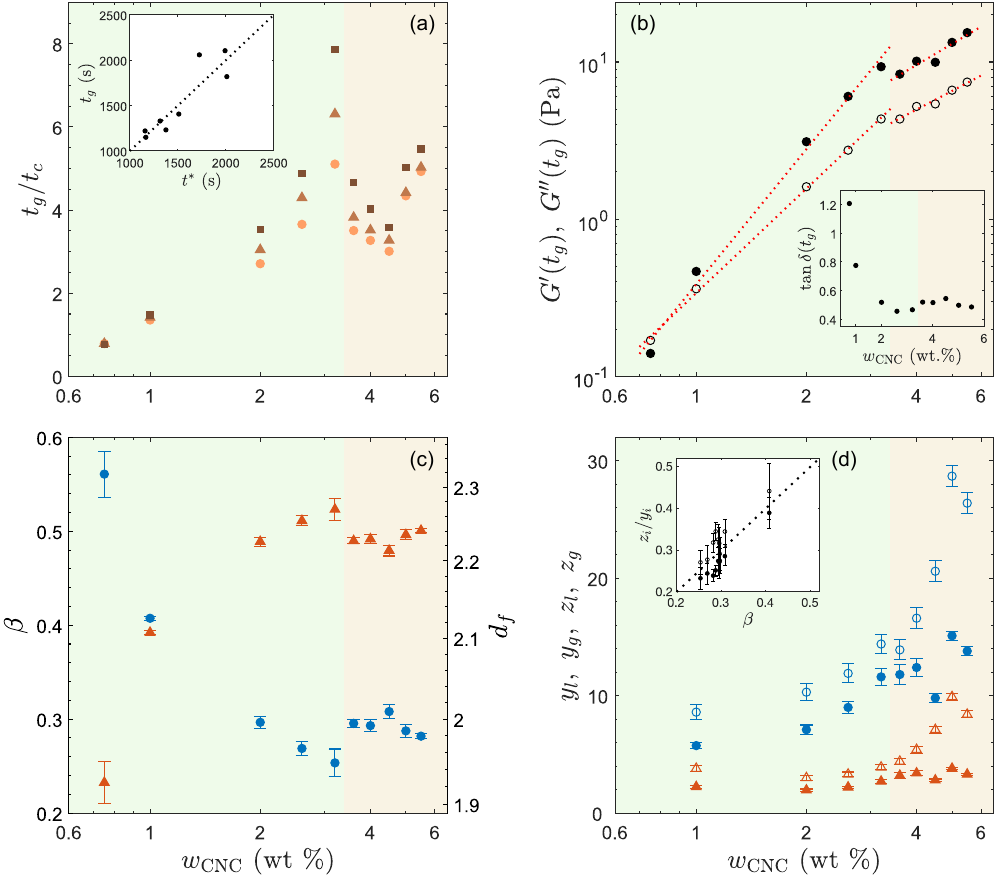}
    \caption{Gelation dynamics parameters as a function of CNC weight fraction $w_\mathrm{CNC}$. Data are shown for sample compositions selected such that the gelation time $t_g$ lies between $1150$~s and $3350$~s (see Table~\ref{tab:parametres_fit} in Appendix~A). (a)~Ratio of the gelation time $t_g$ to the crossover time $t_c$ measured at three different frequencies: $\omega/(2\pi)=0.3$~Hz ($\bullet$), $0.6$~Hz ($\blacktriangle$), and $1.2$~Hz ($\blacksquare$). Inset: Gelation time $t_g$ as a function of the inflection time $t^*$; the dotted line corresponds to $t_g=t^*$. (b)~Elastic modulus $G'(t_g)$ ($\bullet$) and viscous modulus $G''(t_g)$ ($\circ$) measured at the gel point at $\omega/(2\pi)=1.2$~Hz. Red dotted lines indicate power-law fits with exponents $2.9\pm 0.2$ and $1.4\pm 0.1$ for $G'(t_g)$, and $2.2\pm 0.1$ and $1.2\pm 0.1$ for $G''(t_g)$, respectively, below and above $w_\mathrm{CNC}^*= 3.4$~wt\%. Inset: Loss factor at the gel point, $\tan \delta(t_g)$, measured at $1.2$~Hz. (c)~Critical exponent $\beta$ ($\bullet$) and corresponding fractal dimension $d_f$ ($\blacktriangle$). (d)~Power-law exponents characterizing the divergence of the horizontal and vertical shift factors: $y_l$ ($\circ$) and $z_l$ ($\triangle$) in the pre-gel state, and $y_g$ ($\bullet$) and $z_g$ ($\blacktriangle$) in the post-gel state. Inset: Ratio $z_i/y_i$ for $i=l$ ($\circ$) and $i=g$ ($\bullet$) as a function of the critical exponent $\beta$; the dotted line corresponds to $z_l/y_l=z_g/y_g=\beta$. In all graphs, background shading highlight two different regimes as a function of CNC weight fraction: the green region corresponds to a gel phase for $w_\mathrm{CNC}< w_\mathrm{CNC}^*\simeq 3.4$~wt\%, while the orange region corresponds to an attractive glass phase for $w_\mathrm{CNC}>w_\mathrm{CNC}^*$, as discussed in the main text.}
    \label{fig:param_vs_CNC}
\end{figure*}

Finally, the broad relaxation spectra observed throughout gelation motivate a description of the mechanical properties of CNC suspensions in terms of fractional elements. \cite{Keshavarz:2021,Jaishankar:2013,Bonfanti:2020,Bouzid:2018,Bantawa:2023} As already shown for a different composition,\cite{morlet-decarnin:2023} a fractional Maxwell model [see the upper-right sketch in Fig.~\ref{fig:courbes_maitresses_3p2pc_CNC_12mM_NaCl}(b)] captures the master curves in the pre-gel regime, while a fractional Kelvin-Voigt model [see the lower-right sketch in Fig.~\ref{fig:courbes_maitresses_3p2pc_CNC_12mM_NaCl}(c)] describes the post-gel regime [see red curves in Figs.~\ref{fig:courbes_maitresses_3p2pc_CNC_12mM_NaCl}(c) and \ref{fig:courbes_maitresses_3p2pc_CNC_12mM_NaCl}(f)]. The transition between liquid-like and solid-like behavior at the gel point is mediated by a common fractional element ($\mathbb{G}$, $\beta$), or ``spring-pot'', which accounts for the pure power-law viscoelastic response at the gel point.\cite{morlet-decarnin:2023} Fitting the viscoelastic spectra at $t=t_g$ yields an exponent $\beta=0.25$ for the present sample. Once $\beta$ is fixed, each model contains four remaining free parameters, with the constraint that the quasi-property $\mathbb{G}$ remains identical in both models. The resulting parameters determined for this sample are reported in Table~\ref{tab:parametres_fit} in Appendix~A.

Having established the relevance of the time-connectivity superposition on both sides of the gel point for a specific composition ($w_\mathrm{CNC}$=3.2~wt\% and [NaCl]=12~mM), we now proceed to a systematic investigation of the recovery dynamics as a function of CNC weight fraction (Section~\ref{section 4}) and NaCl concentration (Section~\ref{section 5}).

\section{Influence of the CNC weight fraction on the gel recovery dynamics}
\label{section 4}

In this section, we apply the same data processing as described above to CNC suspensions with weight fractions ranging from 0.75~wt\% to 5.5~wt\%. For each CNC content, the NaCl concentration was chosen such that the gelation time $t_g$ is long enough to resolve the recovery dynamics both before and after the gel point, and short enough to avoid solvent evaporation and drying of the sample (see Table~\ref{tab:parametres_fit} in Appendix~A for the corresponding values of $t_g$). The influence of the NaCl concentration at fixed CNC weight fraction will be explored in Section~\ref{section 5}.

\subsection{Robustness of the time-connectivity superposition principle}
\label{time_connectivity_var_CNC}

Figure~\ref{fig:courbes_maitresses_var_CNC_concentration} shows the master curves for the loss factor $\tan \delta(\tilde{\omega})$ and for the rescaled viscoelastic moduli $\tilde{G'}(\tilde{\omega})$ and $\tilde{G''}(\tilde{\omega})]$, together with the corresponding time-dependent shift factors $a(t)$ and $b(t)$, obtained for four different CNC weight fractions, $w_\mathrm{CNC}=1$, 2.6, 4, and 5.5~wt\%. In all cases, a gel point is unambiguously identified, where the loss tangent reaches a plateau [see horizontal dotted lines in Figs.~\ref{fig:courbes_maitresses_var_CNC_concentration}(a)--(d)] and the elastic and viscous moduli follow the same power-law dependence on frequency. All master curves are well fitted by a fractional Maxwell model in the pre-gel regime and by a fractional Kelvin-Voigt model in the post-gel regime (see red curves in Fig.~\ref{fig:courbes_maitresses_var_CNC_concentration} and fit parameters reported in Table~\ref{tab:parametres_fit} in Appendix~A). These results confirm that the time-connectivity superposition principle applies on both sides of the gel point over the whole range of CNC weight fractions explored here.

More quantitatively, the value of the critical exponent $\beta$ is found to vary with the CNC content. In addition, in order to satisfactorily fit the data for CNC weight fractions above 3.2~wt\%, the constraint of imposing the same value for $\mathbb{G}$ on both sides of the gel point has to be relaxed. This observation suggests a change in the nature of the gel for $w_\mathrm{CNC}\gtrsim 3.4$~wt\%, as will be further discussed in Section~\ref{param_tg_var_CNC}. Furthermore, the shift factors $a(t)$ and $b(t)$ systematically display an asymmetric power-law divergence at the gel point (see insets in Fig.~\ref{fig:courbes_maitresses_var_CNC_concentration}), with exponents that depend on the CNC weight fraction. In the following, we examine several observables to characterize the influence of CNC concentration in more detail.

\subsection{Gelation time {\it vs.}~crossover time}
\label{tg_vs_tc_var_CNC}

We first consider the characteristic time scales of the recovery dynamics introduced in Section~\ref{recovery_1HZ}.  
Figure~\ref{fig:param_vs_CNC}(a) shows that the ratio of the frequency-independent gelation time $t_g$ to the frequency-dependent crossover time $t_c$ strongly increases with increasing CNC weight fraction. For $w_\mathrm{CNC}=0.75$~wt\%, the gelation time $t_g$ is slightly shorter than the crossover time $t_c$, whereas for $w_\mathrm{CNC}\ge 1$~wt\%, $t_g$ always exceeds $t_c$, reaching values up to three to eight times larger. For $w_\mathrm{CNC}\gtrsim 3.4$~wt\% the ratio exhibits a more complex, non-monotonic behavior.

Estimating the overlap weight fraction, $w_\mathrm{CNC}^\mathrm{ov}$, by assimilating CNCs to equivalent spheres of diameter equal to their length, in the random close packing limit,\cite{morlet-decarnin:2023} one finds $w_\mathrm{CNC}^\mathrm{ov}\simeq 1$~wt\%. 
This suggests that, relative to the emergence of dominant elastic behavior at 1~Hz signaled by $t_c$, the gelation dynamics governed by $t_g$ are slowed down by steric hindrance as CNCs overlap above $w_\mathrm{CNC}^\mathrm{ov}\simeq 1$, which eventually leads to the decoupling between the two characteristic time scales at large CNC weight fractions, as observed in Section~\ref{recovery_multiwave} for $w_\mathrm{CNC}=3.2$~wt\%.

Finally, the inset of Fig.~\ref{fig:param_vs_CNC}(a) confirms that the inflection time $t^*$ extracted from single-frequency measurements coincides with the gelation time $t_g$ for all CNC weight fractions studied. This result demonstrates that the gel point of CNC suspensions can be reliably identified from the inflection point of $G'(t)$ at a single frequency, rather than from the crossover between $G'(t)$ and $G''(t)$. 

\subsection{Viscoelastic parameters at the gel point}
\label{param_tg_var_CNC}

Figure~\ref{fig:param_vs_CNC}(b) displays the elastic and viscous moduli at the gel point as a function of the CNC weight fraction. For $w_\mathrm{CNC}=0.75$~wt\%, the elastic modulus is smaller than the viscous modulus at the gel point, whereas $G'(t_g)>G''(t_g)$ for $w_\mathrm{CNC}\ge 1$~wt\%. Equivalently, the loss factor $\tan \delta (t_g)$ exceeds unity at the lowest CNC content, but becomes significantly smaller than unity at higher weight fraction [see inset in Fig.~\ref{fig:param_vs_CNC}(b)]. This behavior indicates that, while the gel point consistently marks the onset of mechanical rigidity, the relative contributions of elastic and viscous dissipation at this point depend on the CNC concentration. Below $w_\mathrm{CNC}^\mathrm{ov}\simeq 1$~wt\%, rigidity emerges in a weakly elastic critical state dominated by dissipation, whereas at higher CNC contents, the mechanically percolated network is already strongly constrained, leading to a more elastic response at the gel point.

Moreover, both $G'(t_g)$ and $G''(t_g)$ increase as power laws of $w_\mathrm{CNC}$, with exponents $2.9\pm 0.2$ and $2.2\pm 0.1$, respectively, for $w_\mathrm{CNC}\lesssim 3.4$~wt\% [see red fits in Fig.~\ref{fig:param_vs_CNC}(b)]. Above this threshold, a clear change of regime is observed: both moduli $G'(t_g)$ and $ G''(t_g)$ follow much weaker power-law dependences with exponents between 1.1 and 1.3, resulting in a roughly constant value $\tan \delta (t_g)\simeq 0.5$ at large CNC weight fraction.

\subsection{Critical exponent at the gel point}
\label{beta}

The change of regime at $w_\mathrm{CNC}\simeq 3.4$~wt\%, revealed by the complex behavior in $t_g/t_c$ and by the power-law scaling of $G'(t_g)$ and $G''(t_g)$, and highlighted by green and red background colors in Fig.~\ref{fig:param_vs_CNC}, is also clearly visible in the critical exponent $\beta$ measured at the gel point. As reported in Fig.~\ref{fig:param_vs_CNC}(c), $\beta$ decreases from about 0.55 to 0.25 as CNC $w_\mathrm{CNC}\lesssim 3.4$~wt\%, and then saturates at a value close to 0.3 at larger CNC weight fraction.

Furthermore, assuming that the pure power-law viscoelastic behavior observed at the gel point corresponds to a self-similar microstructure, one may estimate the fractal dimension $d_f$ of the CNC network using the relation proposed by Muthukumar:\cite{Muthukumar:1989} 
\begin{equation}
    d_f=\frac{5}{2} \frac{(3-2\beta)}{(3-\beta)}\,.
    \label{eq.df}
\end{equation}
This relationship, originally derived for polymer gels in the limit of fully screened hydrodynamic and excluded-volume, yields values of $d_f$ increasing from about 1.9 to 2.3 over the range of CNC weight fractions explored here [see right axis in Fig.~\ref{fig:param_vs_CNC}(c)]. These values of $d_f$ are consistent with previous scattering and microscopy measurements reported for suspensions at CNC weight fractions up to 1.5~wt\%.\cite{Moud:2020,phan:2016,Cherhal:2015} Nevertheless, to the best of our knowledge, no measurement of $d_f$ for suspensions with larger CNC weight fractions has been reported yet.

\subsection{Critical exponents for the shift factors $a(t)$ and $b(t)$}
\label{y_z_var_CNC}

Figure~\ref{fig:param_vs_CNC}(d) show that all critical exponents associated with the divergence of the shift factors, namely $y_l$, $y_g$, $z_l$, and $z_g$ increase with $w_\mathrm{CNC}$, making the critical-like divergence of both time-dependent shift factors $a(t)$ and $b(t)$ steeper for denser suspensions. This increase is particularly pronounced for the pre-gel exponents $y_l$ and $z_l$ at high CNC concentrations ($w_\mathrm{CNC}\gtrsim 3.4$~wt\%). Moreover, since $y_i$ is always much larger than $z_i$, the horizontal shift factor $a(t)$ exhibits a stronger time dependence than the modulus shift factor $b(t)$, in agreement with previous studies on other colloidal systems \cite{Suman:2020,he:2021} and with the fact that $z_i/y_i\simeq\beta$ with $\beta<1$, as will be discussed later in Section~\ref{hyp_relations_var_CNC} [see inset of Fig.~\ref{fig:param_vs_CNC}(d)].   

Importantly, the critical exponents differ systematically on either side of the gel point, with $y_l \neq y_g$ and $z_l \neq z_g$ for all CNC weight fractions. This pronounced asymmetry between pre-gel and post-gel dynamics appears to be a distinctive feature of CNC dispersions. In addition, the values of the exponents $y_i$, ranging from 5.7 to 28.7, and $z_i$, ranging from 1.9 to 9.9, are significantly larger than those found in the colloid literature.\cite{Suman:2020}

\subsection{Rheological evidence for a transition from a gel to an attractive glass}
\label{gel_glass_var_CNC}

The recovery dynamics of CNC suspensions following shear cessation reveal a clear separation between two characteristic timescales: the crossover time $t_c$, identified from single-frequency measurements, and the gelation time $t_g$, defined as the time at which the loss tangent becomes independent of frequency. While these two timescales are relatively close at low CNC concentration, their ratio $t_g/t_c$ increases markedly as the CNC weight fraction increases. This separation provides key insight into the nature of the arrested states formed at high concentration.

A particularly striking feature of the data is the existence of a well-defined crossover concentration, $w_\mathrm{CNC}=w_\mathrm{CNC}^* \simeq 3.4$~wt\% at which several rheological observables exhibit a clear change of regime. As shown in Fig.~\ref{fig:param_vs_CNC}(a)--(c), the gelation time, the viscoelastic moduli at the gel point, and the power-law exponent characterizing the gel point all display either a saturation or a significantly weaker dependence on $w_\mathrm{CNC}$ above this concentration. In contrast, below $w_\mathrm{CNC}^*$, these quantities vary strongly with CNC concentration. This change in behavior strongly suggests that the system undergoes a qualitative transition in the nature of the arrested state around this concentration.

To rationalize this observation, it is essential to distinguish between \textit{connectivity} and \textit{mechanical rigidity}. The crossover time $t_c$ corresponds to the emergence of a solid-like response at a given probing frequency, reflecting the formation of connected clusters capable of sustaining stress over a finite timescale. As the CNC concentration increases, particle contacts and connectivity are established increasingly rapidly after shear cessation, leading to smaller values of $t_c$. By contrast, the gelation time $t_g$ corresponds to a stricter mechanical criterion. At $t_g$, the viscoelastic response becomes scale-free, which marks the formation of a \textit{mechanically percolated} network capable of transmitting stress across all probed timescales. In this sense, $t_g$ marks the onset of true mechanical rigidity.

At high CNC concentrations, connectivity is established rapidly after shear cessation, as electrostatic screening by added salt allows negatively charged CNC rods to come into contact and form a space-spanning network. This early connectivity is reflected in the relatively small values of the crossover time $t_c$. However, such a percolated network is not immediately mechanically rigid. Owing to the anisotropic shape of CNCs and their residual electrostatic repulsion, interparticle contacts initially remain mechanically inefficient: they allow for sliding and rotational motion and do not sufficiently constrain the network degrees of freedom.

The establishment of true mechanical rigidity, marked by the gelation time $t_g$, therefore requires slower microstructural rearrangements leading to the formation of mechanically constraining, multi-contact junctions that suppress both translational and rotational modes. As the CNC concentration increases, these rearrangements become increasingly hindered by a combination of steric crowding and residual electrostatic repulsion, which disfavors compact, strongly locked configurations. As a result, mechanical percolation is progressively delayed relative to connectivity percolation, leading to a pronounced increase of the ratio $t_g/t_c$ with $w_\mathrm{CNC}$.

Altogether, these results suggest that increasing the CNC concentration leads to a progressive decoupling between the formation of a connected particle network and the emergence of mechanical rigidity. While electrostatic screening enables rapid contact percolation at high $w_\mathrm{CNC}$, the combined effects of particle anisotropy, steric crowding, and residual electrostatic repulsion delay the formation of a mechanically constraining network. Therefore, we propose that the marked change of regime at $w_\mathrm{CNC} \simeq 3.4$~wt\% reflects a transition from a gel, in which the connectivity and the rigidity develop concomitantly, to an attractive glass-like arrested state, in qualitative agreement with the general phase diagram proposed for CNC suspensions in the presence of salt. \cite{Xu:2020}

\begin{figure}[!t]
    \centering
    \includegraphics[width=0.35\textwidth]{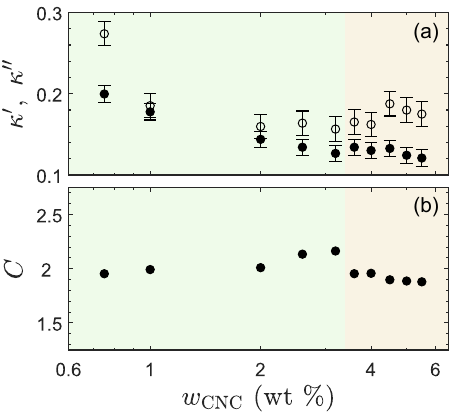}
    \caption{Dynamic critical exponents associated with the gelation dynamics as a function of CNC concentration. (a)~Dynamic critical exponents $\kappa'$ ($\bullet$) and $\kappa''$ ($\circ$) extracted from the frequency dependence of the temporal derivatives of $\ln G'$ and $\ln G''$, respectively, evaluated at the gel point. (b)~Proportionality constant $C$ relating the temporal evolution of $G'$ and $G''$, calculated by imposing $\kappa=\kappa'$ in Eq.~\eqref{eq:kappa} (see main text). Data are shown as a function of the CNC weight fraction $w_\mathrm{CNC}$ and extracted for the same samples as in Fig.~\ref{fig:param_vs_CNC}.}
    \label{fig:param_vs_CNC_kappa}
\end{figure}

\subsection{Dynamic critical exponents}
\label{dyn_crit_exp}

As first introduced by Scanlan and Winter,\cite{Scanlan:1991} one can define another critical exponent characteristic of the gelation dynamics, referred to as the \textit{dynamic} critical exponent $\kappa$, defined through:
\begin{equation}
    \frac{\partial{\ln G'}}{\partial{t}}(\omega,t_g) = C \frac{\partial{\ln G''}}{\partial{t}}(\omega,t_g) \sim \omega^{-\kappa},
    \label{eq:kappa}
\end{equation}
where $C$ is a proportionality factor. This relationship, deduced from experimental observations, suggests that, \corr{at the gel point}, the rates of change of both the elastic and the viscous moduli decrease with the angular frequency following similar power-law behaviors. Based on experimental data obtained on various colloidal systems,\cite{Negi:2014,Suman:2020}, it has been hypothesized that both $\kappa$ and $C$ take universal values, namely $\kappa\simeq 0.2$ and $C\simeq 2$. While the physical interpretation of the value of $\kappa$ remains unclear, $C\simeq 2$ implies that, at the critical gel point, the growth rate of $G'(t)$ is simply about twice that of $G''(t)$.

In practice, we extract two separate exponents $\kappa'$ and $\kappa''$, from the frequency dependence of the time derivatives of $\ln G'$ and $\ln G''$, respectively, evaluated at the gel point. The resulting exponents are plotted in Fig.~\ref{fig:param_vs_CNC_kappa}(a). These two exponents are compatible within experimental uncertainty only for $w_\mathrm{CNC}=1$ and 2~wt\%, and clearly differ in the other cases, in particular in the attractive glass region for $w_\mathrm{CNC}> w_\mathrm{CNC}^*$. 

In order to remain consistent with Eq.~\eqref{eq:kappa} and to estimate the proportionality constant $C$, we therefore impose $\kappa=\kappa'$, which is less noisy than $\kappa''$. As shown in Fig.~\ref{fig:param_vs_CNC_kappa}(b), we find $C\simeq 2.0\pm 0.1$ for all CNC suspensions investigated here, in agreement with previous findings. In contrast, $\kappa$ decreases significantly with increasing $w_\mathrm{CNC}$, from the expected value of 0.2 down to approximately 0.12. Therefore, these results \corr{question} both the hypotheses that $\kappa'=\kappa''$ and that $\kappa=0.2$ should hold independently of the colloidal system under study. 

\subsection{Hyperscaling relations}
\label{hyp_relations_var_CNC}
\begin{figure}[!t]
    \centering
    \includegraphics[width=0.3\textwidth]{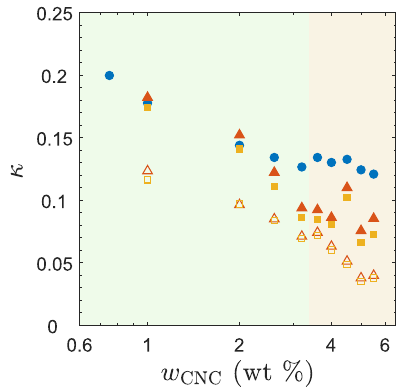}
    \caption{Comparison between experimental and predicted dynamic critical exponents. Dynamic critical exponent $\kappa=\kappa'$ extracted from experimental data ($\bullet$) compared with the values predicted by the different hyperscaling relations: $\kappa=(1-\beta)/(y_l-z_l)$ ($\triangle$), $\kappa=\beta/z_g$ ($\blacktriangle$), $\kappa=1/y_l$ ($\square$), and $\kappa=1/y_g$ ($\blacksquare$), as a function of the CNC weight fraction $w_\mathrm{CNC}$. Data are extracted for the same samples as in Fig.~\ref{fig:param_vs_CNC}.}
    \label{fig:param_vs_CNC_kappa_th_exp_vs_CNC}
\end{figure}

As shown in the literature, the various critical exponents introduced to characterize gelation dynamics around the gel point are linked through mathematical relationships known as \textit{hyperscaling relations}.
\cite{Winter:1987,Stauffer:1982,Scanlan:1991} 
\corr{Following classical notations adapted to the present naming conventions (see Table~\ref{tab:notations} in Appendix~B), in the pre-gel state, the steady-state viscosity $\eta_0$ and the longest relaxation time $\tau_{\mathrm{max},l}$ diverge respectively as $\eta_0\sim\varepsilon^{-s}$ and $\tau_{\mathrm{max},l}\sim\varepsilon^{-1/\kappa_l}$, where $\varepsilon$ denotes the distance to the critical point, i.e., $\varepsilon=|t-t_g|/t_g$ in the case of a time-driven sol-gel transition. A first hyperscaling relation links the exponents $s$ and $\kappa_l$ to the critical exponent $\beta$ through:
\begin{equation}
    \kappa_l=\frac{1-\beta}{s}\,.
    \label{eq:hyperscaling_kappa_l}
\end{equation}
In the post-gel state, the equilibrium elastic modulus $G_0$ increases from zero and the longest relaxation time $\tau_{\mathrm{max},l}$ decreases respectively as $G_0\sim\varepsilon^{z}$ and $\tau_{\mathrm{max},l}\sim\varepsilon^{-1/\kappa_g}$, where the exponents $z$ and $\kappa_g$ are linked to the critical exponent $\beta$ through a second hyperscaling relation:
\begin{equation}
    \kappa_g=\frac{\beta}{z}\,.
    \label{eq:hyperscaling_kappa_g}
\end{equation}
As detailed in Appendix~B, the shift factors $a(t)$ and $b(t)$ respectively correspond to a characteristic time and to the inverse of a characteristic elastic modulus, such that $s\equiv y_l-z_l$, $y_l\equiv 1/\kappa_l$, $z_g\equiv z$, and $y_g\equiv 1/\kappa_g$. Therefore, in the present notations, Eqs.~\eqref{eq:hyperscaling_kappa_l} and \eqref{eq:hyperscaling_kappa_g} simply rewrite as:}
\begin{equation}
    \beta=\frac{z_l}{y_l}=\frac{z_g}{y_g}\,.
    \label{eq:hyperscaling_beta}
\end{equation}
In our previous work,\cite{morlet-decarnin:2023} this relationship was verified for $w_\mathrm{CNC}=2$~wt\% and various salt concentrations. The insets of Figs.~\ref{fig:courbes_maitresses_3p2pc_CNC_12mM_NaCl}(e) and \ref{fig:param_vs_CNC}(d) confirm that the hyperscaling relation given by Eq.~\eqref{eq:hyperscaling_beta} holds for all CNC weight fractions explored here, on both sides of the gel point. \corr{The fact that our experimental data obey hyperscaling relations independently in the pre-gel and post-gel states is an important check that the system consistently passes through a critical gel point.}

\corr{Furthermore, in a very recent theoretical work\cite{Joshi:2026pp}, Joshi shows that the continuity of the time derivatives of $G'$ and $G''$ at the gel point requires that $\kappa_l=\kappa_g=\kappa$, where $\kappa$ is the dynamical exponent defined above in Sect.~\ref{dyn_crit_exp}, such that:}
\begin{equation}
    \kappa=\frac{1}{y_l} \mbox{~in the pre-gel state,}\label{eq:hyperscaling_kappa_p_3}
\end{equation}
and
\begin{equation}
    \kappa=\frac{1}{y_g} \mbox{~in the post-gel state.} \label{eq:hyperscaling_kappa_p_4}    
\end{equation}
\corr{In other words, from a theoretical standpoint, the gelation dynamics should display a symmetric divergence of the longest relaxation time, i.e., $y_l=y_g$. According to Eq.~\eqref{eq:hyperscaling_beta}, this should also automatically lead to the symmetry of the moduli dynamics, $z_l=z_g$. Under such a symmetry condition, Eqs.~\eqref{eq:hyperscaling_kappa_l} and \eqref{eq:hyperscaling_kappa_g} can be respectively rewritten as:}
\begin{equation}
    \kappa=\frac{1-\beta}{y_l-z_l} \mbox{~in the pre-gel state,}
    \label{eq:hyperscaling_kappa_p_1}
\end{equation}
and
\begin{equation}
    \kappa=\frac{\beta}{z_g} \mbox{~in the post-gel state.}
    \label{eq:hyperscaling_kappa_p_2}
\end{equation}

Figure~\ref{fig:param_vs_CNC_kappa_th_exp_vs_CNC} compares the experimentally extracted dynamic critical exponent $\kappa=\kappa'$ already shown in Fig.~\ref{fig:param_vs_CNC_kappa}(a) to the values predicted by the four hyperscaling relations given by Eqs.~\eqref{eq:hyperscaling_kappa_p_3}--\eqref{eq:hyperscaling_kappa_p_2}. The two predictions derived from the pre-gel dynamics [Eqs.~\eqref{eq:hyperscaling_kappa_p_3} and \eqref{eq:hyperscaling_kappa_p_1}] coincide --see empty symbols in Fig.~\ref{fig:param_vs_CNC_kappa_th_exp_vs_CNC}-- as do those derived from the post-gel dynamics [Eqs.~\eqref{eq:hyperscaling_kappa_p_4}) and \eqref{eq:hyperscaling_kappa_p_2}] --see filled symbols in Fig.~\ref{fig:param_vs_CNC_kappa_th_exp_vs_CNC}). This confirms that the hyperscaling relation given by Eq.~\eqref{eq:hyperscaling_beta} holds in our system for all CNC weight fractions. 
\begin{figure}[!t]
    \centering
    \includegraphics[width=0.35\textwidth]{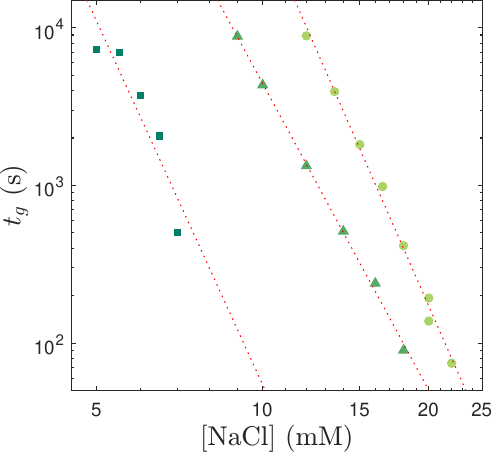}
\caption{Dependence of the gelation time on salt concentration. Gelation time $t_g$ as a function of NaCl concentration for three series of CNC suspensions with different CNC weight fractions (see Table~\ref{tab:parametres_fit_multiwave_var_NaCl} in Appendix~A: $w_\mathrm{CNC}=2$~wt\% ($\bullet$), 3.2~wt\% ($\blacktriangle$), and 5.5~wt\% ($\blacksquare$). Red dashed lines correspond to the best power-law fits at fixed CNC weight fraction, $t_g\sim \mbox{[NaCl]}^{-\nu}$, yielding exponents $\nu=8.0$, $6.5$, and $7.7$ from right to left.}
    \label{fig:param_var_CNC_vs_NaCl_tg}
\end{figure}
\corr{Yet}, consistent with the pronounced asymmetry between pre-gel and post-gel dynamics in the present CNC suspensions, the two sets of predictions, from $t<t_g$ and from $t>t_g$, differ significantly. Interestingly, the exponents predicted from the post-gel dynamics are in good agreement with the experimental values of $\kappa'$ at low CNC weight fraction ($w_\mathrm{CNC}<3$~wt\%). By contrast, all hyperscaling relations fail to predict the dynamic critical exponent in the attractive glass region of the phase diagram, indicating a breakdown of critical gelation scaling as dense, attraction-dominated arrest is approached.

\begin{figure*}[!t]
    \centering
    \includegraphics[width=0.85\textwidth]{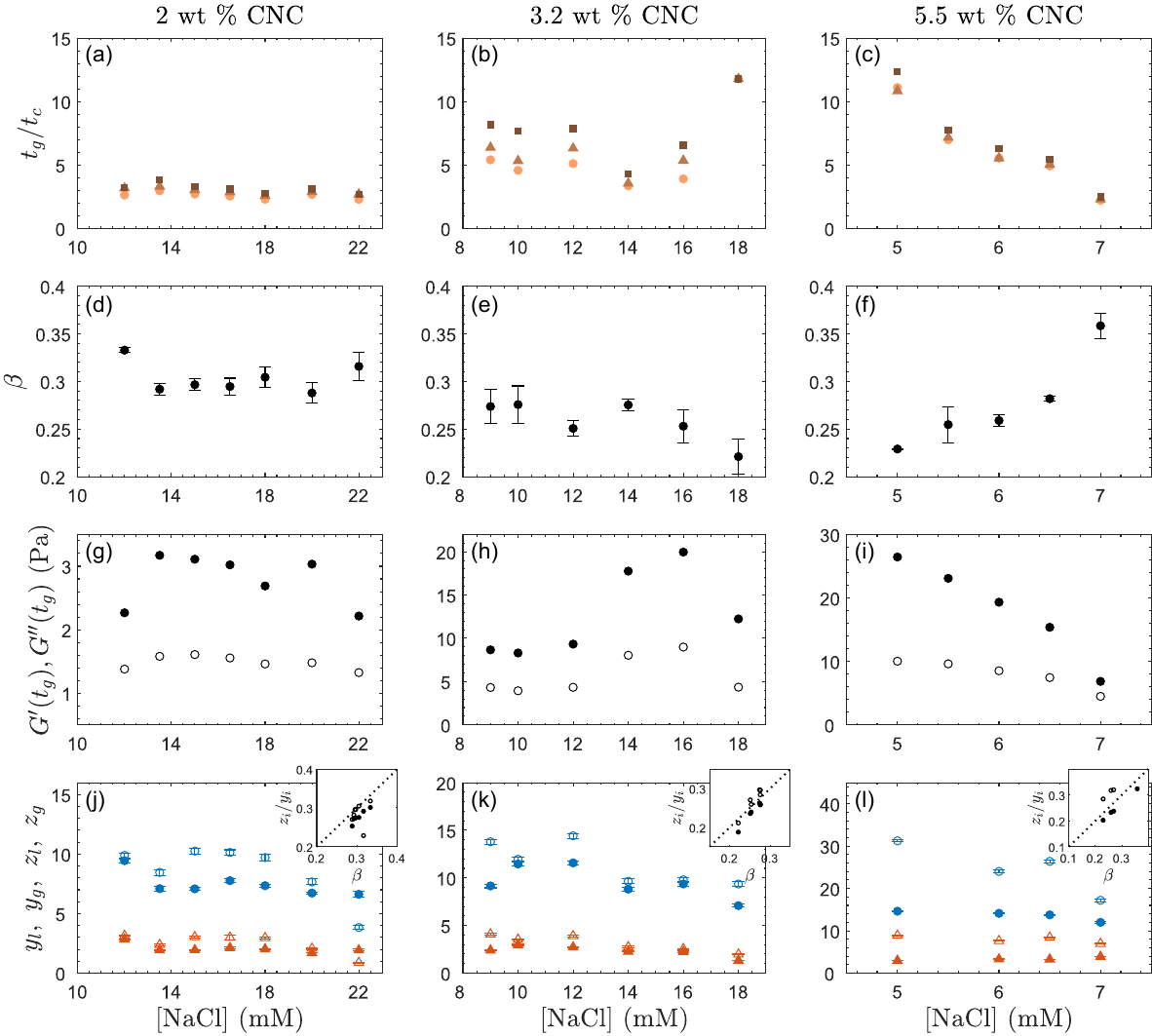}
    \caption{Effect of salt concentration on gelation dynamics parameters. Gelation dynamics parameters as a function of [NaCl] for three series of suspensions with different CNC weight fractions (see Table~\ref{tab:parametres_fit_multiwave_var_NaCl} in Appendix~A): $w_\mathrm{CNC}=2$~wt\% (left column), 3.2~wt\% (middle column), and 5.5~wt\% (right column). (a)--(c)~Ratio of the gelation time $t_g$ to the crossover time $t_c$ measured at three different frequencies: $\omega/(2\pi)=0.3$~Hz ($\bullet$), 0.6~Hz ($\blacktriangle$), and 1.2~Hz ($\blacksquare$). (d)--(f)~Critical exponent $\beta$. (g)-(i)~Elastic modulus $G'(t_g)$ ($\bullet$) and viscous modulus $G''(t_g)$ ($\circ$) measured at the gel point at $\omega/(2\pi) =1.2$~Hz. (j)--(l)~Power-law exponents characterizing the divergence of the horizontal and vertical shift factors, $y_l$ ($\circ$) and $z_l$ ($\triangle$) in the pre-gel state, and $y_g$ ($\bullet$) and $z_g$ ($\blacktriangle$) in the post-gel state. Insets: Ratio $z_i/y_i$ for $i=l$ ($\circ$) and $i=g$ ($\bullet$) as a function of the critical exponent $\beta$; the dotted line corresponds to $z_l/y_l=z_g/y_g=\beta$.}
    \label{fig:param_var_CNC_vs_NaCl}
\end{figure*}

\section{Influence of the ionic strength on the gel recovery dynamics}
\label{section 5}

Having explored the influence of the CNC content on the recovery dynamics of samples with similar gelation times, we now investigate the effect of ionic strength by applying time-resolved mechanical spectroscopy to samples containing various NaCl concentrations at fixed CNC content. We focus on three series of samples containing $w_\mathrm{CNC}=2$~wt\%, 3.2~wt\%, and 5.5~wt\%, each prepared with 5 to 7 different NaCl concentrations, leading to gelation times between 70~s and 9000~s. 
According to the discussion of Section~\ref{gel_glass_var_CNC}, these CNC weight fractions fall respectively in the gel phase, close to the gel-attractive glass boundary, and in the attractive glass phase.

\begin{figure*}[!t]
    \centering
    \includegraphics[width=0.85\textwidth]{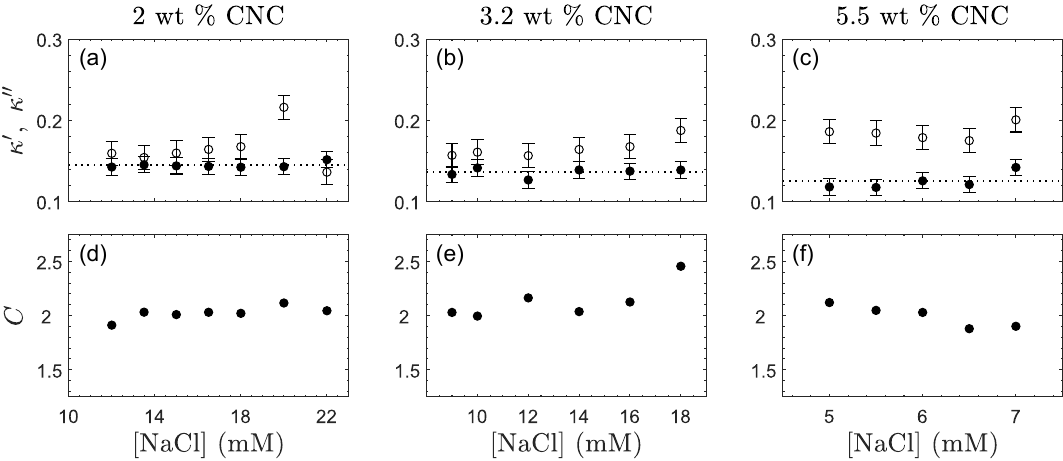}
    \caption{Dynamic critical exponent and proportionality constant as a function of salt concentration. (a)--(c) Dynamic critical exponents $\kappa'$ ($\bullet$) and $\kappa''$ ($\circ$) extracted from the temporal evolution of $G'$ and $G''$, respectively, and (d)--(f) proportionality constant $C$ calculated by setting $\kappa$ equal to the mean value of $\kappa'$ over the explored range of NaCl concentrations [see horizontal gray dashed lines in (a)--(c)], plotted as a function of NaCl concentration. Data are extracted for the same samples as in Fig.~\ref{fig:param_var_CNC_vs_NaCl}.}
    \label{fig:param_var_CNC_vs_NaCl_kappas_C}
\end{figure*}

In all cases, master curves similar to those shown in Figs.~\ref{fig:courbes_maitresses_3p2pc_CNC_12mM_NaCl} and \ref{fig:courbes_maitresses_var_CNC_concentration} can be constructed using the same rescaling analysis as in previous sections (see also Fig.~3 in Ref.~\onlinecite{morlet-decarnin:2023} for the series with $w_\mathrm{CNC}=2$~wt\%). This confirms the robustness of the time-connectivity superposition principle over the whole range of NaCl and CNC concentrations under study. Furthermore, the same fitting procedure based on fractional models yields the parameters reported in Table~\ref{tab:parametres_fit_multiwave_var_NaCl} in Appendix~A. As observed previously when varying the CNC content, the constraint of imposing the same prefactor $\mathbb{G}$ on both sides of the gel point must again be relaxed for the largest CNC content $w_{\mathrm{CNC}}=5.5$~wt\%. In the following, we characterize in more detail the dependence of the gel recovery dynamics on salt concentration using the various observables introduced above in Section~\ref{section 4}.

\subsection{Dependence of the gelation kinetics on the salt content}
\label{tg_var_salt}

Figure~\ref{fig:param_var_CNC_vs_NaCl_tg} shows the gelation times $t_g$ as a function of NaCl concentration for the three CNC contents investigated.
As expected, the gelation kinetics accelerates sharply as the screening of the CNC surface charges becomes more efficient with increasing [NaCl].
Interestingly, for all three series the gelation time follows a power-law dependence on salt concentration, $t_g \sim \textrm{[NaCl]}^{-\nu}$, with exponents $\nu=8.0$, $6.5$, and $7.7$, for $w_\mathrm{CNC}=2$~wt\%, $3.2$~wt\%, and $5.5$~wt\%, respectively. Similar values have been reported for CNC suspensions in the presence of NaCl: in a previous work focusing on the inflection time $t^*$, we reported $\nu=8.7$ for $w_\mathrm{CNC}=3.2$~wt\% CNC and $\textrm{[NaCl]}=5$--100~mM,\cite{Morlet-Decarnin:2022} while Peddireddy \textit{et al.} \cite{Peddireddy:2016} reported $\nu=10$ for $w_\mathrm{CNC}= 0.5$~wt\% CNC and $\textrm{[NaCl]}=30$--70~mM.  

More generally, power laws have been observed for the dependence of the gelation time with salt concentration in other colloidal systems, such as silver iodide particles\cite{Reerink:1954} and silica particles,\cite{Linden:2015} with exponents $\nu$ ranging between $5.9$ and $11$ depending on salt nature and particle concentration. Theoretical approaches developed for monodisperse spherical particles predict similarly large exponents, together with a significant influence of counter-ion valency.\cite{Reerink:1954,Linden:2015,Zaccone:2014} The absence of such a dependence in CNC suspensions\cite{Morlet-Decarnin:2022} might be due to the strong geometrical anisotropy of the CNCs and the dominant role of directional and steric constraints during aggregation.

\subsection{Rheological observables at the gel point}
\label{gel_point_var_salt}

As shown in Fig.~\ref{fig:param_var_CNC_vs_NaCl}(a,d,g), for $w_\mathrm{CNC}=2$~wt\% the rheological observables characterizing the gel point are nearly independent of NaCl concentration, with $t_g/t_c\simeq 2.9\pm 0.4$, $\beta\simeq 0.30\pm 0.02$, $G'(t_g)\simeq 2.8\pm 0.4$~Pa, and $G''(t_g)\simeq 1.5\pm 0.1$~Pa. In contrast, for $w_\mathrm{CNC}=5.5$~wt\% [see Fig.~\ref{fig:param_var_CNC_vs_NaCl}(c,f,i)], all these quantities display a pronounced dependence on [NaCl]: $t_g/t_c$ decreases from about $12$ to $2$, $\beta$ increases from $0.23$ to $0.36$, and both $G'(t_g)$ and $G''(t_g)$ decrease substantially over a narrow salt concentration range.

Such markedly different behaviors confirm that these two CNC contents belong to distinct regions of the phase diagram associated with different gelation scenarios. As discussed in Section~\ref{gel_glass_var_CNC}, in the gel phase ($w_\mathrm{CNC}<w_\mathrm{CNC}^*\simeq 3.4$~wt\%), CNCs are initially well separated in the dispersed state. Upon recovery, they must translate over distances comparable to their length in order to aggregate and form a mechanically percolating network. In this regime, salt concentration primarily controls aggregation kinetics, while the network microstructure at the gel point remains largely insensitive to [NaCl]. 

By contrast, for $w_\mathrm{CNC}>w_\mathrm{CNC}^*$, CNCs are sufficiently close that percolation can be achieved primarily through rotational rearrangements. In this attractive glass phase, electrostatic screening plays a more direct structural role: increasing [NaCl] facilitates faster sticking between neighboring rods, promoting the formation of a more open, less rigid structure at the gel point. This is reflected in the decrease of the fractal dimension, or equivalently, to the increase of the exponent $\beta$ [see Fig.~\ref{fig:param_var_CNC_vs_NaCl}(f)], and the concomitant decrease of the elastic modulus $G'(t_g)$  [see Fig.~\ref{fig:param_var_CNC_vs_NaCl}(i)] as [NaCl] increases. 

Finally, the intermediate case $w_\mathrm{CNC}=3 .2$~wt\% exhibits mixed behavior [see Fig.~\ref{fig:param_var_CNC_vs_NaCl}(b,e,h)]. Rheological observables remain nearly constant at low salt content but display complex, non-monotonic variations for $\textrm{[NaCl]}\gtrsim 12$~mM. This behavior likely reflects the proximity of the gel-attractive glass boundary and suggests that the position of the boundary itself may also depend on the salt content.

\subsection{Critical exponents and hyperscaling relations}
\label{hyp_rel_var_salt}

As shown in Fig.~\ref{fig:param_var_CNC_vs_NaCl}(j-l) and Fig.~\ref{fig:param_var_CNC_vs_NaCl_kappas_C}(a-c), all critical exponents characterizing the recovery dynamics appear to be independent of salt concentration for the three CNC contents investigated. Moreover, $\kappa'\simeq \kappa''$ for $w_\mathrm{CNC}=2$~wt\% and 3.2~wt\%, while $\kappa'$ is significantly smaller than $\kappa''$ for $w_\mathrm{CNC}=5.5$~wt\%, consistent with Fig.~\ref{fig:param_vs_CNC_kappa}(a). 

Using the mean value of $\kappa'$ for each CNC weight fraction [see horizontal dashed lines in Fig.~\ref{fig:param_var_CNC_vs_NaCl_kappas_C}(a-c)], we estimate the proportionality constant $C$ entering Eq.~\eqref{eq:kappa}. As shown in Fig.~\ref{fig:param_var_CNC_vs_NaCl_kappas_C}(d-f), $C$ is independent of salt concentration and close to $2$ for all CNC contents, in agreement with the hypothesis of Suman and Joshi.\cite{Suman:2020} Nevertheless, $\kappa'$ remains systematically smaller than $0.2$, reinforcing the conclusion that the dynamic critical exponent is not universal, as already suggested in Section~\ref{hyp_relations_var_CNC}.

\begin{figure*}[!t]
    \centering
    \includegraphics[width=0.85\textwidth]{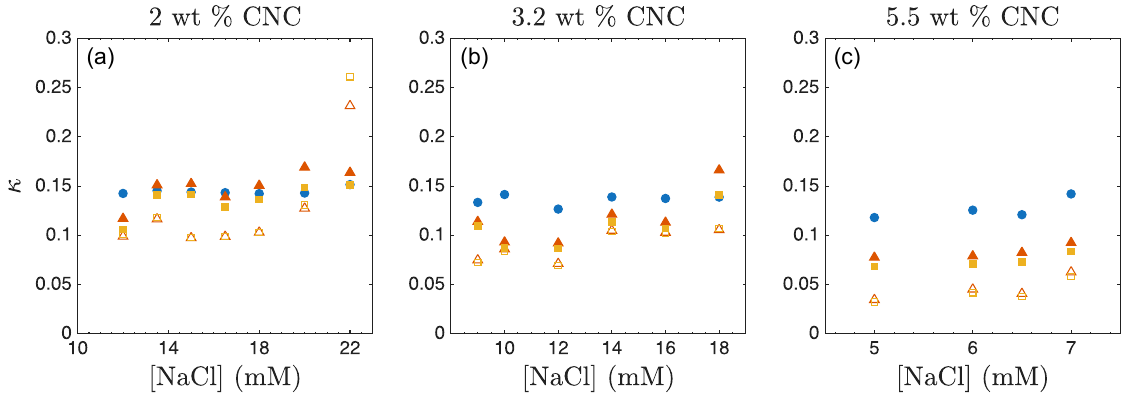}
    \caption{Comparison between experimental and predicted dynamic critical exponents as a function of salt concentration. Dynamic critical exponent $\kappa=\kappa'$ extracted from experimental data ($\bullet$) compared to the values predicted by the different hyperscaling relations: $\kappa=(1-\beta)/(y_l-z_l)$ ($\triangle$), $\kappa=\beta/z_g$ ($\blacktriangle$), $\kappa=1/y_l$ ($\square$), and $\kappa=1/y_g$ ($\blacksquare$), plotted as a function of NaCl concentration. Data are extracted for the same samples as in Fig.~\ref{fig:param_var_CNC_vs_NaCl}.}\label{fig:param_var_CNC_vs_NaCl_kappa_th_exp_vs_NaCl}
\end{figure*}

The insets of Fig.~\ref{fig:param_var_CNC_vs_NaCl}(j-l) further indicate that the hyperscaling relation given by Eq.~\eqref{eq:hyperscaling_beta}
holds for all salt concentrations, both in the pre-gel and post-gel regimes. Finally, Fig.~\ref{fig:param_var_CNC_vs_NaCl_kappa_th_exp_vs_NaCl} compares the experimental values of the dynamic critical exponent with the hyperscaling predictions given by Eqs.~\eqref{eq:hyperscaling_kappa_p_1}-\eqref{eq:hyperscaling_kappa_p_4} and extends the conclusions of Section~\ref{hyp_relations_var_CNC}: predictions based on post-gel dynamics successfully describe the data in the gel phase [see Fig.~\ref{fig:param_var_CNC_vs_NaCl_kappa_th_exp_vs_NaCl}(a)], but fail in the attractive glass phase [see Fig.~\ref{fig:param_var_CNC_vs_NaCl_kappa_th_exp_vs_NaCl}(c)]. For the intermediate CNC weight fraction $w_\mathrm{CNC}=3.2$~wt\%, agreement is recovered only at large salt concentrations [see Fig.~\ref{fig:param_var_CNC_vs_NaCl_kappa_th_exp_vs_NaCl}(b)], where electrostatic screening likely shifts the system deeper into the gel-like regime.

\section{Open questions and conclusion}
\label{section 6}

We have investigated how CNC suspensions recover their solid-like properties in the presence of salt after full fluidization by a strong shear. Using time-resolved mechanical spectroscopy, we have shown that recovery always proceeds through a critical gel point and obeys a time-connectivity superposition principle, as reported for various polymeric and colloidal systems. Yet, our results reveal several striking features that depart from established gelation scenarios and raise a number of open questions.

\subsection{How may one rationalize the difference between the crossover time and the gelation time?}
\label{discussion_tc_vs_tg}

A particularly unexpected result is the pronounced decoupling between the crossover time $t_c$ and the gelation time $t_g$, as soon as the CNC weight fraction is larger than 1~wt\%. In this regime, $t_g$ can exceed $t_c$ by more than an order of magnitude. To the best of our knowledge, previous studies on colloidal gels --including those made of anisotropic particles-- generally reported $t_g\simeq t_c$ when both times were measured.

This decoupling suggests that the onset of measurable elasticity ($G'>G''$ at finite frequency) precedes the establishment of a self-similar, percolated network responsible for critical gel behavior. Steric constraints associated with the rod-like shape of CNCs, their polydispersity, and their tendency to form bundles may delay the emergence of this network. Importantly, CNCs carry a high surface charge, and electrostatic repulsion may further hinder local rearrangements even after mechanical connectivity is achieved and despite the presence of salt. The strong sensitivity of $t_g/t_c$ to salt concentration in the attractive glass phase supports this interpretation, as electrostatic screening directly modulates the ability of densely packed CNCs to reorganize.

\subsection{What is the origin of the asymmetry between pre-gel and post-gel critical dynamics?}
\label{discussion_y_z}

By constructing master curves for the temporal evolution of the viscoelastic spectra across the sol-gel transition, we have shown that the recovery dynamics of CNC suspensions are intrinsically asymmetric with respect to the gel point, with distinct critical exponents before and after percolation. To the best of our knowledge, such asymmetry has only rarely been reported, notably in gelatin systems, where the pre-gel exponent for $a(t)$ was reported to be three times larger than the post-gel exponent.\cite{sun:2018} 

Another specificity of CNC suspensions is that the values of the critical exponents $y_l$, $y_g$, $z_l$, and $z_g$ are significantly larger than those reported for polymeric and colloidal gels. Indeed, the largest value found in the literature for $y_l$ and $y_g$, which are always larger than $z_l$ and $z_g$, is $6.2$ for a suspension of polydisperse hydrogenated castor oil colloidal rods \cite{wehrman:2016}, while $y_l$ and $y_g$ range between 6 and 31 in the present CNC suspensions. Theories have been developed to predict the critical exponents for polymer gels.\cite{Adolf:1990,Winter:1997,Larsen:2008,Winter:1987,Suman:2020,deGennes:1976,Adam:1981,Stauffer:1982,Axelos:1990,Hodgson:1990,Rouwhorst:2020} For instance, the Flory-Stockmayer theory predicts an exponent $z_l=z_g=3$, and the percolation theory predicts a value of 1.7 in three dimensions.\cite{Stauffer:1982} The much larger values reported for CNC suspensions suggest a much larger sensitivity to the proximity of the sol-gel transition, both for $a(t)$, which reflects the divergence of the longest relaxation time in the system as clusters grow,  and for $b(t)$, which reflects the emergence of the elastic modulus in the limit of zero frequency. However, no \corr{quantitative} theory has yet been established \corr{to predict the values of critical exponents in} colloidal gels, even in the case of monodisperse attractive spheres. Therefore, once again, it remains to be seen whether the unexpected asymmetry of the recovery dynamics and the uncommonly large values of the critical exponents $y_l$, $y_g$, $z_l$, and $z_g$ result from the strong shape anisotropy of CNCs, from their size distribution, or from combined effects of electrostatic interactions, and collective rearrangements constrained by crowding.

Nevertheless, the hyperscaling relation $\beta=z_l/y_l=z_g/y_g$ holds separately for both the pre-gel and the post-gel dynamics, reflecting the continuous evolution of the viscoelastic spectra around the gel point, and ensuring a constant value of the critical exponent $\beta$ on both sides of the gel point. \corr{Still, following Ref.~\onlinecite{Joshi:2026pp}, the observed asymmetry of the critical exponents contradicts the continuity of the time derivatives of the viscoelastic moduli across the sol-gel transition. As suggested by Joshi, it remains to be seen whether such a contradiction stems from experimental issues, including uncertainties in the position of the gel point and/or in the time and moduli shift factors, or from more fundamental issues specific to non-equilibrium phenomena in anisotropic systems close to glassy structural arrest. In view of the extensiveness and robustness of the present data set, the latter seems more likely, and we believe that it may be necessary to question the hypotheses of the percolation theory, specifically those that entail the continuity of the time derivatives of $G'$ and $G''$, as well as its applicability to systems such as CNC gels.}

\corr{Finally,} we have interpreted the variations of $\beta$ with CNC and salt concentrations in terms of the fractal dimension $d_f$ of the CNC network at the gel point, by assuming that Eq.~\eqref{eq.df} applies. However, this relationship --originally established for branched polymers \cite{Winter:1997,ng:2008}-- may not strictly hold for a system of rod-like colloids such as CNCs. Structural measurements, e.g., using static light scattering or small-angle X-ray scattering, are therefore required to check the validity of this assumption.

\subsection{How does the gel--attractive glass boundary compares to the literature?}
\label{discussion_gel_glass}

In their review of the literature, Xu \textit{et al.} \cite{Xu:2020} proposed a gel-attractive glass boundary as a continuation of the liquid crystal-repulsive glass boundary (see Figs.~12 and 15 in Ref.~\onlinecite{Xu:2020}), without relying on any experimental data points, and stating that the distinction between the gel and the attractive glass cannot be made through rheology but only through microstructural characterization. In contrast, we have identified a clear transition in the behavior of rheological observables at the gel point, which we associate with a CNC weight fraction $w_\mathrm{CNC}^*\simeq 3.4$~wt\%. Interestingly, computing the corresponding dimensionless concentration $c_\mathrm{glass}=\phi r$, with $\phi$ the CNC volume fraction and $r=L/D$ the CNC aspect ratio, where $L\simeq 120$~nm is the length of an individual CNC and $d\simeq 10$~nm is its diameter, we find $c_\mathrm{glass}\simeq 0.26$, which is significantly lower than the value of $0.7$ proposed in Ref.~\onlinecite{Xu:2020}. Such a discrepancy stands out as an open issue and clearly deserves more investigation.

Finally, structural characterization is essential to confirm the proposed phase boundary between gel and attractive glass regions, and clarify the respective roles of electrostatic screening, steric constraints, and particle anisotropy. \corr{Combining rheology with light, X-ray, or neutron scattering, or with} conductivity measurements probing electrical percolation would provide valuable insight into the microscopic mechanisms underlying gelation and glassy arrest in CNC suspensions. In this context, \corr{X-ray photon correlation spectroscopy has been used very recently to follow the aging dynamics of chemically linked, spherical nanocrystals \cite{Ofosu:2026}, and may constitute an excellent tool to also explore the gelation dynamics of anisotropic colloids such as the present CNCs. Moreover, low-field NMR spectroscopy of the solvent has emerged} as a highly sensitive and complementary probe of network formation and mechanical rigidity,\cite{Legrand:2024,Herveou:2025} even in the absence of long-range structural order, and could offer a powerful way to bridge microscopic constraints and macroscopic rheological response. 

\section*{Author contributions}
LMD: conceptualization, data curation, formal analysis, investigation, methodology, writing -- original draft; TD and SM: conceptualization, formal analysis, funding acquisition, methodology, project administration, resources, supervision, validation, writing -- review \& editing.

\section*{Data availability}
Data corresponding to all figures in this paper are available through the following DOI: 10.6084/m9.figshare.31628899.

\section*{Conflicts of interest}
There are no conflicts to declare.

\section*{Acknowledgements}
The authors thank Dr. Bruno Jean and Dr. Jean-Luc Puteaux (CERMAV, Grenoble) for the TEM measurements shown in Fig.~\ref{fig:TEM}. They also acknowledge fruitful discussions with Dr. Isabelle Capron, Prof. Yogesh Joshi, and Dr. Yu Ogawa. For the purpose of Open Access, a CC-BY 4.0 public copyright licence has been applied by the authors to the present document and will be applied to all subsequent versions up to the Author Accepted Manuscript arising
from this submission.


\section*{Appendix~A: Parameters for fractional models}

Tables~\ref{tab:parametres_fit} and \ref{tab:parametres_fit_multiwave_var_NaCl} gather the parameters inferred from fitting the master curves for viscoelastic moduli with fractional Maxwell and Kelvin-Voigt models as detailed in Section~\ref{time_connectivity}. 

\section*{\corr{Appendix B: Notations for critical exponents}}
\corr{In the present work, we probe the dynamic scaling of the sol-gel transition based on the construction of master curves inferred from time-resolved mechanical spectroscopy. In particular, we introduce two shift factors $a(t)$ and $b(t)$ such that, with $\tilde{\omega}=a(t) \omega$, $\tilde{G'}=b(t)G'$, and $\tilde{G''}= b(t)G''$, the rescaled viscoelastic moduli $\tilde{G'}$ and $\tilde{G''}$ as a function of $\tilde{\omega}$ both become independent of the distance to the gel point $\varepsilon=|t-t_g|/t_g$. Based on dimensional arguments, it can be argued that $a(t)$ is a time that follows the evolution of the system across the sol-gel transition, which must be proportional to the longest relaxation time of the system, $\tau_{\mathrm{max}}$, on both sides of the gel point. Similarly, $b(t)$ is inversely proportional to a characteristic elastic modulus, and must be inversely proportional to the equilibrium modulus $G_0$ in the post-gel state and to $\eta_0/\tau_{\mathrm{max}}$ in the pre-gel state, where $\eta_0$ is the steady-state viscosity.}

\corr{The need for a reference state in the present scaling approach does not allow for a quantitative comparison with the characteristic relaxation times defined by Winter in Ref.~\onlinecite{Winter:1987} from the intersections of the low-frequency asymptote of the (dimensional) dynamic viscosity $|\eta^*|=\sqrt{G'^2+G''^2}/\omega$ and the high-frequency asymptote of $|\eta^*|$ at the gel point. Still, the equivalence between both approaches was shown and discussed in Ref.~\onlinecite{Suman:2020}. Such equivalence leads to:
\begin{equation}
    a\sim\tau_{\mathrm{max},l} \mbox{~and~} b\sim\frac{\tau_{\mathrm{max},l}}{\eta_0} \mbox{~in the pre-gel state,}\label{eq:equiv_pregel}
\end{equation}
and
\begin{equation}
    a\sim\tau_{\mathrm{max},g} \mbox{~and~} b\sim\frac{1}{G_0} \mbox{~in the post-gel state.}\label{eq:equiv_postgel}
\end{equation}
Using the exponents defined in the main text for $a$, $b$, $\tau_{\mathrm{max}}$, $\eta_0$, and $G_0$, this leads to: 
\begin{equation}
    \varepsilon^{-y_l}\sim\varepsilon^{-1/\kappa_l} \mbox{~and~} \varepsilon^{-z_l}\sim\varepsilon^{s-1/\kappa_l}\mbox{~in the pre-gel state,}\label{eq:equiv_pregel}
\end{equation}
and
\begin{equation}
    \varepsilon^{-y_g}\sim\varepsilon^{-1/\kappa_g}  \mbox{~and~} \varepsilon^{-z_g}\sim\varepsilon^{-z} \mbox{~in the post-gel state.}\label{eq:equiv_postgel}
\end{equation}
We may therefore identify $y_l\equiv 1/\kappa_l$, $s\equiv y_l-z_l$, $y_g\equiv 1/\kappa_g$, and $z_g\equiv z$. Table~\ref{tab:notations} provides a full comparison of the various notations used in Refs.~\onlinecite{Winter:1987} and ~\onlinecite{Joshi:2026pp} and in the present work.}

\begin{table*}[h]
    \centering
\begin{tabular}{| c | c  | c | c | c | c | c | c | c | c | c | }
    \hline
      Sample composition & $t_g$ (s) & $\eta$ & $\mathbb{V}$ & $\alpha$ & $\mathbb{G}$ ($t<t_g$) & $\mathbb{G}$ ($t>t_g$) & $\beta$ & $\mathbb{K}$ & $\xi$ & $k$ \\[8pt] 
    \hline
     0.75~wt\% CNC  20~mM NaCl & 1309.5 & N.A. & N.A. & N.A. & \multicolumn{2}{c|}{N.A.} & N.A. & N.A. & N.A. & N.A. \\[8pt]
    \hline
     1~wt\% CNC  17~mM NaCl & 3334.7 & 2.0 & 0.025 & 0.87 & \multicolumn{2}{c|}{0.035} & 0.41 & 0.17 & 0.31 & 0.14 \\[8pt]
      \hline
      2~wt\% CNC  15~mM NaCl & 1818.6 & 0.13 & 0.14 & 0.63 & \multicolumn{2}{c|}{0.95} & 0.30 & 5.4 & 0.23 & 17 \\[8pt]
      \hline
      2.6~wt\% CNC  14~mM NaCl & 1220.5 & 0.27 & 0.23 & 0.55 & \multicolumn{2}{c|}{2.5} & 0.27 & 9.0 & 0.17 & 21 \\[8pt]
      \hline
      3.2~wt\% CNC  12~mM NaCl & 1329.8 & 0.69 & 0.51 & 0.53 & \multicolumn{2}{c|}{3.5} & 0.25 & 25.7 & 0.17 & 29.41 \\[8pt]
      \hline
      3.6~wt\% CNC  11~mM NaCl & 1406.3 & 1.0 & 0.35 & 0.63 & 2.6 & 6.4 & 0.30 & 37.5 & 0.18 & 42.3 \\[8pt]
      \hline
      4~wt\% CNC  10~mM NaCl & 1150.5 & 8.7 & 0.36 & 0.63 & 0.82 & 2.4 & 0.29 & 9.13 & 0.17 & 2.31 \\[8pt]
      \hline
      4.5~wt\% CNC  9~mM NaCl & 1231.9 & 9.0 & 0.15 & 0.67 & 5.5 & 37.5 & 0.31 & 798 & 0.18 & 2318 \\[8pt]
      \hline
      5~wt\% CNC  7~mM NaCl & 2104.2 & 10 & 0.15 & 0.65 & 6.1 & 160 & 0.29 & 9500 & 0.18 & 1.35$\times10^5$ \\[8pt]
      \hline
      5.5~wt\% CNC, 6.5~mM NaCl & 2058.5 & 1.2 & 0.30 & 0.60 & 11 & 113 & 0.27 & 5422 & 0.16 & 2.51$\times10^4$ \\[8pt]
      \hline
\end{tabular}
    \caption{Parameters of the fractional Maxwell and Kelvin-Voigt models used to fit the master curves for the loss tangent $\tan \delta$ and for the viscoelastic moduli $G'$ and $G''$ obtained from the temporal evolution of the viscoelastic spectra during the recovery after strong shear of aqueous CNC suspensions with various weight fractions $w_\mathrm{CNC}$ and NaCl concentrations. For $w_\mathrm{CNC}<3.4$~wt\%, the same value for $\mathbb{G}$ is used on both sides of the gel point.}
    \label{tab:parametres_fit}
\end{table*}

\begin{table*}[h]
    \centering
\begin{tabular}{| c | c  | c | c | c | c | c | c | c | c | c | }
   \hline
      Sample composition & $t_g$ (s) & $\eta$ & $\mathbb{V}$ & $\alpha$ & $\mathbb{G}$ ($t<t_g$) & $\mathbb{G}$ ($t>t_g$) & $\beta$ & $\mathbb{K}$ & $\xi$ & $k$ \\[8pt] 
    \hline
2~wt\% CNC  12~mM NaCl & 8888 & 0.17 & 0.13 & 0.68 & \multicolumn{2}{c|}{0.54} & 0.33 & 2.40 & 0.20 & 3.00
\\[8pt] \hline
    2~wt\% CNC  13.5~mM NaCl & 3939 & 1.00 & 0.19 & 0.60 & \multicolumn{2}{c|}{0.35} & 0.29 & 0.60 & 0.18 & 0.49
\\[8pt] \hline
    2~wt\% CNC  15~mM NaCl & 1818 & 0.13 & 0.14 & 0.63 & \multicolumn{2}{c|}{0.95} & 0.30 & 5.40 & 0.23 & 17.00
\\[8pt] \hline
    2~wt\% CNC  16.5~mM NaCl & 985 & 0.15 & 0.14 & 0.60 & \multicolumn{2}{c|}{1.10} & 0.29 & 3.00 & 0.20 & 8.00
\\[8pt] \hline
    2~wt\% CNC   18~mM NaCl & 415 & 0.23 & 0.14 & 0.63 & \multicolumn{2}{c|}{1.21} & 0.30 & 5.90 & 0.19 & 12.00
\\[8pt] \hline
    2~wt\% CNC  20~mM NaCl & 193 & 0.52 & 0.34 & 0.59 & \multicolumn{2}{c|}{0.85} & 0.29 & 2.90 & 0.18 & 4.60
\\[8pt] \hline
    2~wt\% CNC  22~mM NaCl & 75 & 0.52 & 11410 & 0.70 & \multicolumn{2}{c|}{0.40} & 0.32 & 1.10 & 0.20 & 0.32
\\[8pt] \hline
    3.2~wt\% CNC   9~mM NaCl & 8824 & 0.44 & 0.34 & 0.58 & \multicolumn{2}{c|}{3.79} & 0.28 & 16.80 & 0.24 & 300.00
\\[8pt] \hline
    3.2~wt\% CNC  10~mM NaCl & 4334 & 1.00 & 0.40 & 0.58 & \multicolumn{2}{c|}{4.50} & 0.30 & 0.16 & 0.19 & 38.11
\\[8pt] \hline
    3.2~wt\% CNC  12~mM NaCl & 1330 & 0.69 & 0.51 & 0.53 & \multicolumn{2}{c|}{3.50} & 0.25 & 25.70 & 0.17 & 29.41
\\[8pt] \hline
    3.2~wt\% CNC  14~mM NaCl & 512 & 0.67 & 1.14 & 0.47 & \multicolumn{2}{c|}{22.40} & 0.23 & 9.50 & 0.12 & 7.13
\\[8pt] \hline
    3.2~wt\% CNC  16~mM NaCl & 239 & 2.40 & 2.03 & 0.46 & \multicolumn{2}{c|}{17.40} & 0.23 & 9.95 & 0.13 & 6.50
\\[8pt] \hline
    3.2~wt\% CNC  18~mM NaCl & 90 & 2.50 & 2.27 & 0.48 & \multicolumn{2}{c|}{3.63} & 0.23 & 19.50 & 0.13 & 20.00
\\[8pt] \hline
      5.5~wt\% CNC  5~mM NaCl & 7306 & 0.28 & 0.64 & 0.51 & 189 & 189 & 0.23 & $3.03 \times 10^4$ & 0.18 & $2.58 \times 10^6$
\\[8pt] \hline
        5.5~wt\% CNC  5.5~mM NaCl & 6963 & N.A. & N.A. & N.A. & N.A. & N.A. & N.A. & N.A. & N.A. & N.A.
\\[8pt] \hline
        5.5~wt\% CNC  6~mM NaCl & 3735 & 0.33 & 0.32 & 0.57 & 14.50 & 95.80 & 0.26 & $1.93 \times 10^3$ & 0.15 & $4.00 \times 10^3$
\\[8pt] \hline
        5.5~wt\% CNC  6.5~mM NaCl & 2059 & 1.20 & 0.30 & 0.60 & 11.00 & 113.00 & 0.27 & 5422 & 0.16 & 2.51$\times10^4$
\\[8pt] \hline
       5.5~wt\% CNC  7~mM NaCl & 505 & $5.79 \times 10^3$ & 2.00 & 0.78 & 0.52 & 1.90 & 0.36 & 5.21 & 0.17 & 0.2
\\[8pt] \hline
  \end{tabular}
  \caption{Parameters of the fractional Maxwell and Kelvin-Voigt models used to fit the master curves for the loss tangent $\tan \delta$ and for the viscoelastic moduli $G'$ and $G''$ obtained from the temporal evolution of the viscoelastic spectra during the recovery after strong shear of aqueous CNC suspensions with $w_\mathrm{CNC}=2$~wt\%, 3.2~wt\%, and 5.5~wt\% and various NaCl concentrations. For $w_\mathrm{CNC}=2$~wt~\% and $w_\mathrm{CNC}=3.2$~wt~\%, the same value for $\mathbb{G}$ is used on both sides of the gel point.}    \label{tab:parametres_fit_multiwave_var_NaCl}
\end{table*}

\begin{table*}[h]
    \centering
    \includegraphics[width=0.9\textwidth]{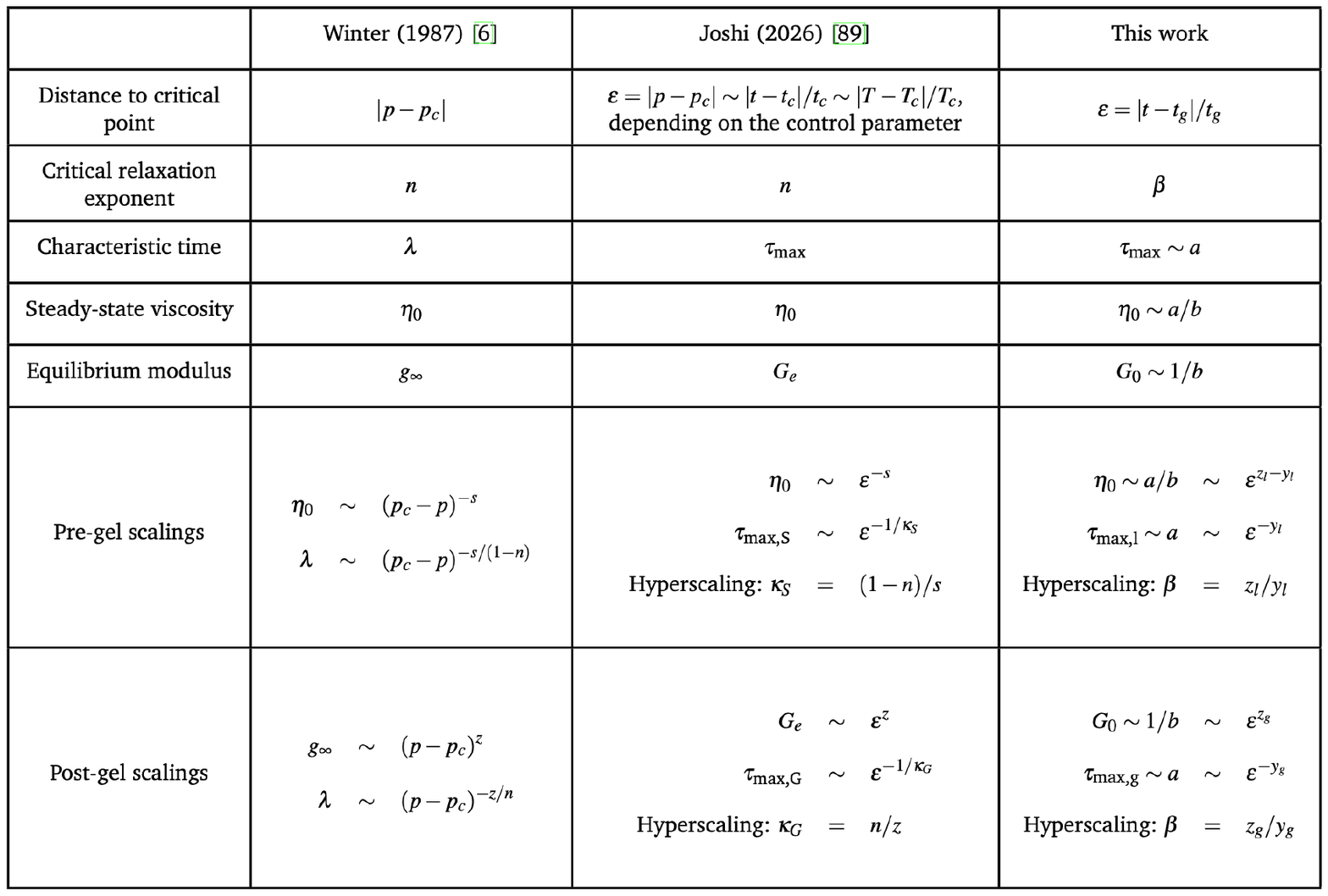}

    \caption{\corr{Notations used in Refs.~\onlinecite{Winter:1987} and \onlinecite{Joshi:2026pp} compared to those used in the present work. $p$ is the degree of cross-linking and $p_c$ its critical value, $t$ is the time and $t_c$ or $t_g$ the gelation time, and $T$ is the temperature and $T_c$ its critical value. All exponents are assumed to be positive numbers.}}   
    \label{tab:notations}
\end{table*}

\end{document}